\documentclass[%
 superscriptaddress,
 preprint,
 amsmath,
 amssymb,
 prl,
 showpacs,
 floatfix,
 nofootinbib,
]{revtex4-1}

\usepackage{graphicx}
\usepackage{dcolumn}
\usepackage{bm}
\usepackage[colorlinks=true, allcolors=blue]{hyperref}
\usepackage{color}
\usepackage{braket}
\usepackage{url}
\usepackage{xcolor}
\usepackage{soul}
\usepackage{lineno}

\usepackage[utf8]{inputenc}
\DeclareUnicodeCharacter{2212}{-}

\usepackage[T1]{fontenc}
\usepackage[scaled=0.92]{helvet}

\usepackage{caption}
\usepackage[justification=raggedright,singlelinecheck=false]{caption}

\begin{document}

\title{Irradiation-induced amplification of electric fields at oxide interfaces as revealed by correlative DPC-STEM and DFT}

\author{Elizabeth A. Peterson} \email{epeterson@lanl.gov} \affiliation{Theoretical Division, Los Alamos National Laboratory, Los Alamos, NM 87545, USA}
\author{Dongye Liu} \affiliation{Materials Science and Engineering, University of California Berkeley, Berkeley, CA 94720, USA}
\author{Sean H. Mills} \email{shm925@lehigh.edu} \affiliation{Materials Science and Engineering, University of California Berkeley, Berkeley, CA 94720, USA}\affiliation{Materials Science and Engineering, Lehigh University, Bethlehem, PA 18015, USA}
\author{Tiffany C. Kaspar}\affiliation{Physical and Computational Sciences Directorate, Pacific Northwest National Laboratory, Richland, WA 99354, USA}
\author{Hyosim Kim}\affiliation{Materials Science and Technology Division, Los Alamos National Laboratory, Los Alamos, NM 87545, USA}
\author{Yongqiang Wang}\affiliation{Materials Science and Technology Division, Los Alamos National Laboratory, Los Alamos, NM 87545, USA}
\author{Blas P. Uberuaga}\affiliation{Materials Science and Technology Division, Los Alamos National Laboratory, Los Alamos, NM 87545, USA}
\author{Andrew M. Minor} \affiliation{Materials Science and Engineering, University of California Berkeley, Berkeley, CA 94720, USA} \affiliation{National Center for Electron Microscopy, Molecular Foundry, Lawrence Berkeley National Laboratory, Berkeley, CA 94720, USA}

\begin{abstract}
Heterointerfaces are ubiquitous in modern devices, found in technologies ranging from microelectronics to structural components for energy applications. Many of these emerging technologies are found in applications such as satellites, batteries, and next generation nuclear reactors, that  are subject to harsh environments. In some scenarios, multiple extreme conditions, such as irradiation and corrosion, act on the material simultaneously. Extending the lifetime of these technologies is dependent on a detailed understanding of how their component materials platforms and  interfaces respond in extreme environments, where irradiation and corrosion may couple in unique ways, distinct from corrosion under ambient conditions. Oxides, which form readily over metal underlayers, can act as protective coatings; enhancing the robustness of oxide overlayers to protect underlying metal alloys is a potential avenue towards corrosion mitigation. Here we study the impact of irradiation-induced non-equilibrium defects on charge segregation and electric fields at and near multi-phase oxide heterointerfaces. We perform a detailed study of irradiated Fe$_{2}$O$_{3}$-Cr$_{2}$O$_{3}$ thin film heterostructures using first-principles density functional theory (DFT) electronic structure modeling paired with four-dimensional-scanning transmission electron microscopy (4D-STEM) differential phase contrast (DPC) and electron energy loss spectroscopy (EELS) techniques to measure nanoscale changes in electric fields. Our results show clear evidence that irradiation drives substantial modulation of interfacial electric fields that can be tailored by controlling the atomistic chemical structure of the oxide interface. We show that irradiation can selectively induce built-in electric fields, thereby altering their direction; this suggests a pathway to engineering protective oxide heterostructure overlayers that can electrically control the spatial distribution of defects, with significant implications for the design of corrosion-resistant materials for extreme environments.

\end{abstract}

\maketitle

\section{\label{sec:intro}Introduction}

The presence of heterointerfaces often dictates the generation, evolution, and transport of defects, electronic carrier behavior, and chemistry of multicomponent materials. In non-metals, this interfacial influence can arise, at least in part, from the electrostatic effects at the interface. In semiconductors, these electrostatic effects manifest via the band alignment, chemical potential, and interfacial dipoles, all of which may lead to the presence of internal electric fields. Understanding these electrostatic effects is crucial for predicting physical behavior of heterointerfaces in applications ranging from microelectronic devices to the corrosion of structural alloys. An illustration of the influence of interfacial electrostatic effects on physical properties is provided by epitaxial heterostructures of Fe$_{2}$O$_{3}$ and Cr$_{2}$O$_{3}$. In prior work, we showed that these heterostructures exhibited interfacial electronic band offsets that were highly dependent on the order of growth \cite{Chambers2000, Jaffe2004, Kaspar2016}. Growth order was observed to affect the atomistic details of the interfacial elemental structure, leading to chemistry-dependent electrostatic differences. These electrostatic differences give rise to non-commutative values of the band offset that depend on the growth order, and we showed that superlattices of Fe$_{2}$O$_{3}$ and Cr$_{2}$O$_{3}$ can utilize these differing band offsets to induce a built-in potential across the superlattice structure of up to 0.8 eV \cite{Kaspar2016}. 

Despite the central role of interfacial electric fields on material behavior, few probes exist to characterize them in detail. Internal electric fields are typically studied with area-averaged techniques such as X-ray photoelectron spectroscopy (XPS) \cite{Kaspar2016} or surface measurements such as atomic force microscopy (AFM) \cite{Song2022}. These techniques have limited ability to probe deeply buried interfaces and do not offer simultaneous spatial and chemical information. Recent developments in four-dimensional scanning transmission electron microscopy (4D-STEM) differential phase contrast (DPC) imaging have enabled precise mapping of internal electric fields with nanometer-scale resolution. This technique leverages a pixelated detector to record a full 2D diffraction pattern at each scan position, allowing local electric fields to be quantified by tracking shifts in the center of mass (CoM) of the diffraction disks, which are directly proportional to the projected electric field gradient \cite{Ophus2023}. Compared to traditional DPC using segmented detectors, 4D-STEM DPC offers substantially improved spatial resolution ($\leq$ 1 nm), quantitative sensitivity to weak fields ($<$ 0.1 MV/cm), and the ability to cover large fields of view ($>$ 1 $\mathrm{\mu m^{2}}$) \cite{Chen2024, daSilva2022}. Moreover, recent advances in compressive sensing \cite{Smith2025} and deep learning-based noise reduction \cite{Lee2022} have significantly extended the applicability of this method to beam-sensitive oxides, enabling real-time tracking of defect-related field changes with minimal sample damage.

Oxide heterointerfaces are ubiquitous, both intentionally manufactured as in microchips or unintentionally evolved via oxidation of device components. A representative example is the case of corrosion of Fe-Cr steel, which can result in segregation into distinct binary oxide regions; typically these consist of Cr-rich inner oxides and Fe-rich outer oxides such as Fe$_{3}$O$_{4}$ / Cr$_{2}$O$_{3}$ ~\cite{Kruska2012}. Due to this ubiquity, understanding how oxide interfaces respond to harsh environments will be critical for the development of many emerging technologies, with applications ranging from microelectronics, to space technologies, to nuclear reactors. As a specific example, next generation nuclear reactors may be built from steels or other transition metal alloys that will be subject to simultaneous irradiation and corrosion ~\cite{Samin2019a, Samin2019b, Schmidt2021, Haseman2021, Lach2021, Agarwal2022}. In such extreme environments, irradiation and corrosion may couple in unique ways, resulting in outcomes that are distinct from corrosion mechanisms under ambient conditions ~\cite{Agarwal2020, Schmidt2021, Haseman2021, Yano2021a, Yano2021b, Lach2021, Yano2022, OwusuMensah2022, Chan2023, Xi2024, LopezMorales2024, Liu2025, Sooby2025, Zhao2025}. Irradiation of oxides and oxide heterostructures is expected to produce  non-equilibrium populations of a wide variety of charged point defects, including ion vacancies and interstitials. First-principles modeling of these defects typically focuses on the thermodynamics and kinetics of defect generation and migration ~\cite{Samin2019a, Samin2019b, Li2020, Banerjee2020, Banerjee2021, Schmidt2021, Yano2021a, Yano2021b, Yano2022, OwusuMensah2022, Agarwal2022, Chan2023, Banerjee2023, Hatton2023, Xi2024, Liu2025}, but these studies typically do not consider the effects of electronic structure and electrostatics on these charged defects. This poses the question of whether migration of charged point defects can be influenced by built-in electrostatic effects at oxide interfaces. Should oxide heterostructure interfaces be engineered to achieve the selective electrical sequestration of defects within a single oxide, this capability would yield substantial progress in developing corrosion-resistant materials for extreme environments.

In this work, we characterize the behavior of charged defects at Fe$_{2}$O$_{3}$-Cr$_{2}$O$_{3}$ interfaces under ion irradiation by coupling \textit{ab initio} density functional theory (DFT) defect electronic structure modeling with nanoscale electric field measurement via 4D-STEM DPC imaging. We investigate whether built-in electric fields play a significant role in interfacial stability and compositional evolution post-irradiation. We further elucidate the interplay between interfacial atomistic structure and irradiation and their implications for corrosion. The results from these advanced simulation and characterization techniques provide a new pathway for predicting material degradation in harsh environments by elucidating how irradiation alters built-in electric fields and defect distributions at oxide interfaces.

\section{\label{sec:results}Results}

\begin{figure}[h]
    \centering
    \includegraphics[width=0.8\linewidth]{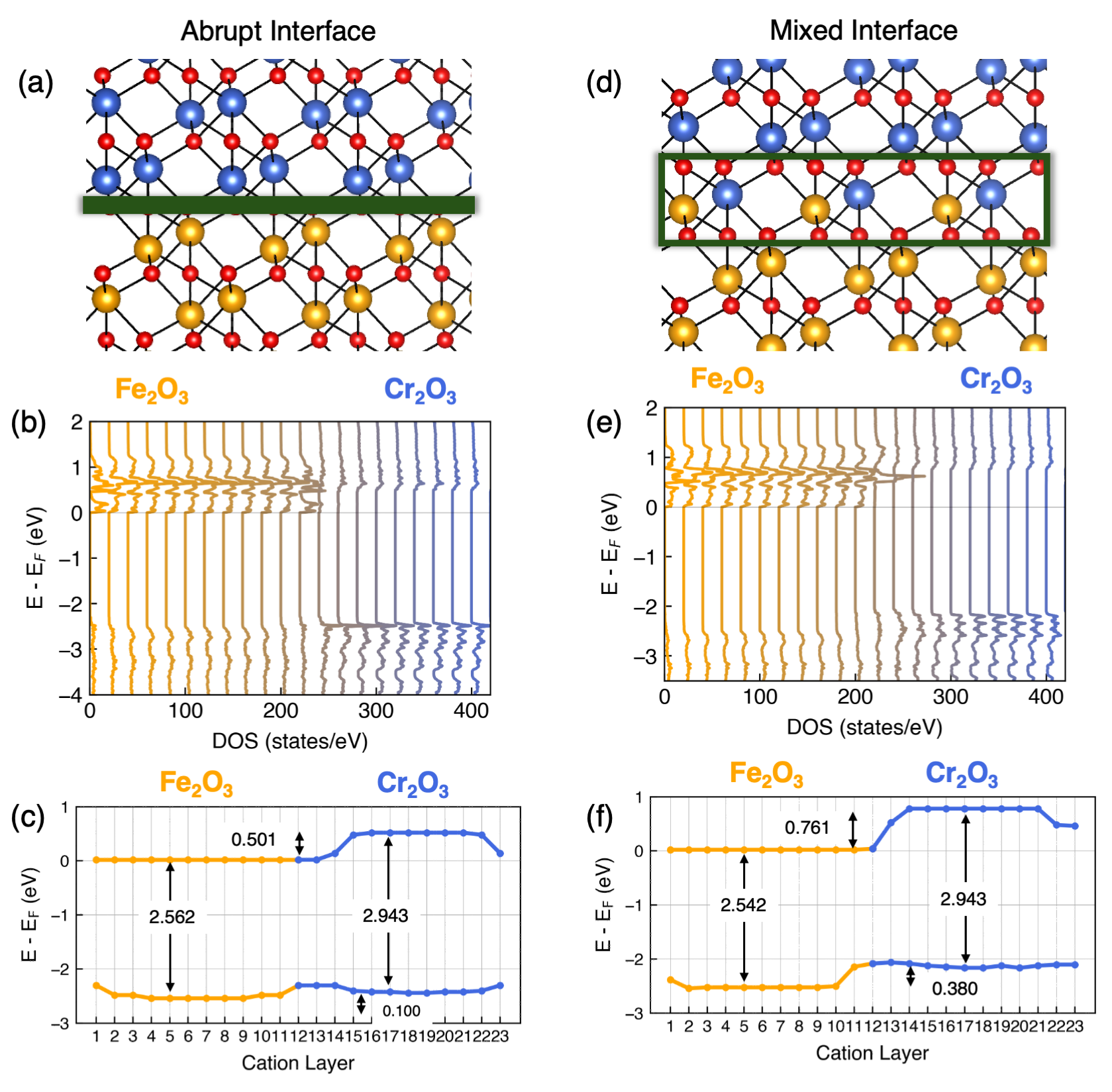}
    \caption{\textbf{The atomistic chemical and structural details of each interface type and associated layer-resolved density of states.} For the (a) abrupt interface, each cation layer only contains one type of cation, Fe (gold) or Cr (blue), while for the (d) mixed interface the interfacial cation layer contains both Fe and Cr. Oxygen ions are shown in red. The layer-resolved density of states and relative band edge positions of the (b,c) abrupt interface and (e,f) mixed interface.}
    \label{fig:strucs_dos}
\end{figure}

In the corundum lattice along the (0001) direction, a layered structure of cation bilayers (two slightly offset layers of cations) are separated by layers of oxygen ions. When Fe$_{2}$O$_{3}$ is grown on top of Cr$_{2}$O$_{3}$, we previously reported ~\cite{Kaspar2016} that the interface structure had an abrupt transition where a bilayer of Fe atoms is followed by a bilayer of Cr atoms. However, when Cr$_{2}$O$_{3}$ is grown on top of Fe$_{2}$O$_{3}$, the interface is characterized by a mixed bilayer of both Fe and Cr atoms. These two interfacial structures are illustrated in Fig. \ref{fig:strucs_dos}(a),(d).

To probe the built-in electric fields associated with atomistic chemical microstructure, we calculated the DFT electronic structure of interfacial heterostructures of Fe$_{2}$O$_{3}$-Cr$_{2}$O$_{3}$ with both abrupt (Fig. \ref{fig:strucs_dos}(a)) and mixed interfaces (Fig. \ref{fig:strucs_dos}(d)) using the DFT+U method of Dudarev, et al. ~\cite{Dudarev1998}.

The density of states (DOS) for abrupt and mixed interfaces with 12 layers of each oxide ($2\times2\times2$ supercells with 24 cation layers total) was calculated and shows a quantifiable difference in band offset between the Fe$_{2}$O$_{3}$ and Cr$_{2}$O$_{3}$ regions depending on the atomistic structure of the interface. In both cases, the bulk band edges extracted from projected DOS calculations are shifted to higher energy for Cr$_{2}$O$_{3}$  than those of Fe$_{2}$O$_{3}$ (Fig. \ref{fig:strucs_dos}(b,e)), indicating a built in electric field with varying magnitude but oriented in the same direction for each type of interface. The bulk band gaps in each region are consistent irrespective of interface type, at 2.5 eV for Fe$_{2}$O$_{3}$ and 2.9 eV for Cr$_{2}$O$_{3}$. These values are in reasonable agreement with experimentally measured values of 2.0-2.2 eV ~\cite{Mishra2015} and 3.0-3.2 eV ~\cite{Cao2006,Abdullah2014}, respectively, indicating that our supercells were large enough to capture representative bulk-like oxide regions. The band offset for the abrupt interface (Fig. \ref{fig:strucs_dos}(b),(c)) is $\sim$0.25 eV smaller than for the mixed interface (Fig. \ref{fig:strucs_dos}(e),(f)). The valence band maximum (VBM) offset is 0.10 eV for the abrupt interface and 0.38 eV for the mixed interface. Similarly, the conduction band minimum (CBM) offset is 0.50 eV for the abrupt interface and 0.76 eV for the mixed interface. Prior experimental results measured the VBM offset from Fe$_{2}$O$_{3}$ to Cr$_{2}$O$_{3}$ in a mixed interface as +0.8 eV; the measured VBM offset from Cr$_{2}$O$_{3}$ to Fe$_{2}$O$_{3}$ in an abrupt interface was -0.4 eV ~\cite{Kaspar2016}. The trend, namely that the VBM offset is larger for the mixed interface than for the abrupt interface, is in qualitative agreement with our calculated values of 0.380 eV from Fe$_{2}$O$_{3}$ to Cr$_{2}$O$_{3}$ in the mixed interface and -0.10 eV from Cr$_{2}$O$_{3}$ to Fe$_{2}$O$_{3}$ in the abrupt interface. The quantitative discrepancy can be attributed to limitations of the DFT exchange-correlation method employed.

Band offset is a proxy for the electrostatic thermodynamic barrier for charged carriers and defects to migrate from one oxide to the other. The distinct electrostatic barriers to charge transfer in the two pristine interfaces and the presence of built-in electric fields of varying magnitudes suggests that migration of defects, especially charged defects, should be affected by the interfacial atomistic structure with implications for corrosion and coupled irradiation-corrosion mechanisms and propensities.

\begin{figure}
    \centering
    \includegraphics[width=1.0\linewidth]{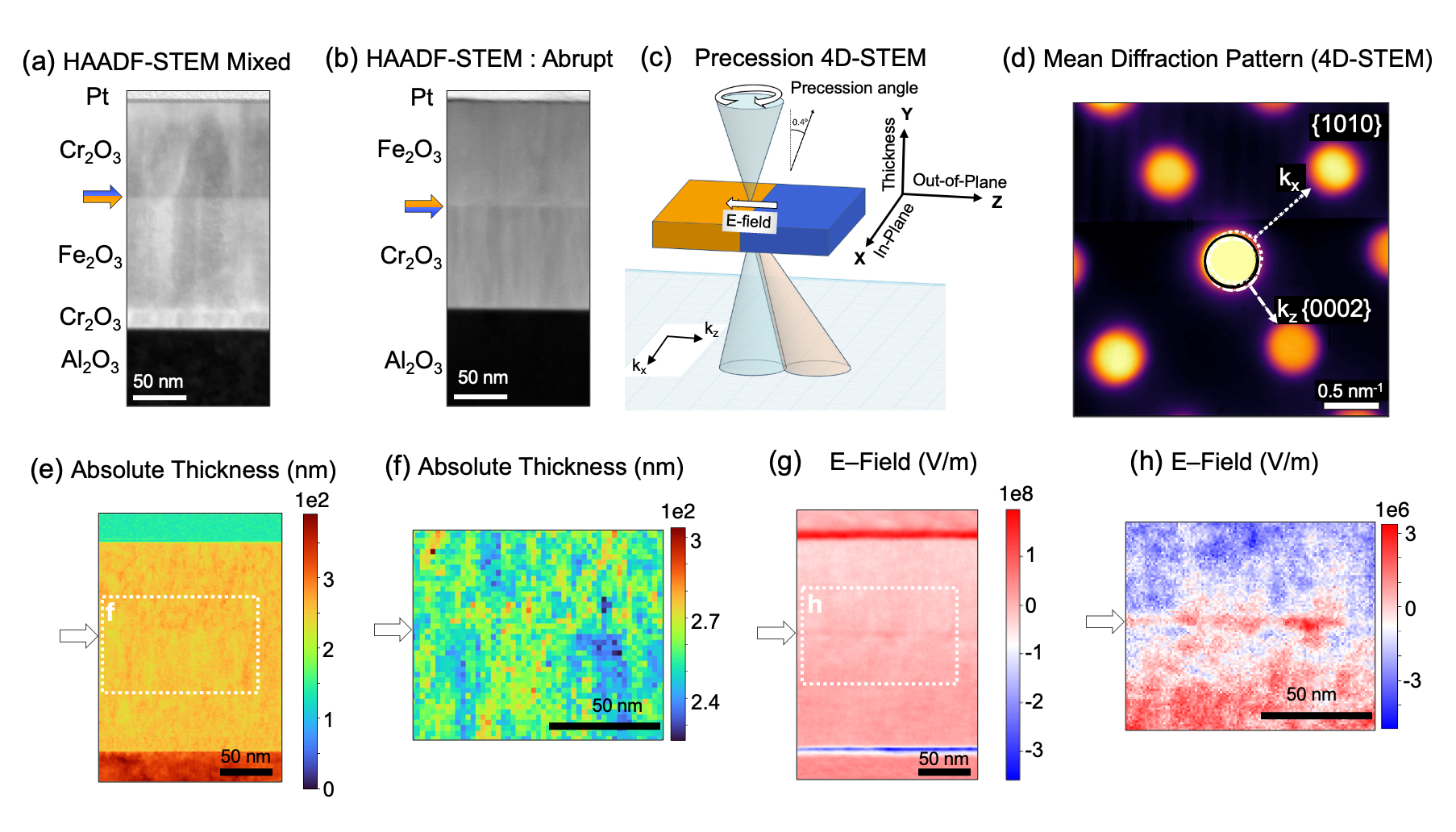}
    \caption{\textbf{The setup and selected data for the 4D-STEM and DPC-PED measurements.} HAADF image of (a) Cr$_{2}$O$_{3}$ / Fe$_{2}$O$_{3}$ mixed interface and (b) Fe$_{2}$O$_{3}$ / Cr$_{2}$O$_{3}$ abrupt interface after irradiation. (c) Schematic of the DPC-PED setup for oxide interfaces with a nearly parallel beam illumination. k$_{x}$ and k$_{z}$ are the x-axis and z-axis in reciprocal space when collecting data and calculating the CoM shifts. (d) Mean diffraction pattern of 4D-STEM dataset from irradiated Fe$_{2}$O$_{3}$ / Cr$_{2}$O$_{3}$  with an abrupt interface. Fe$_{2}$O$_{3}$ and Cr$_{2}$O$_{3}$ are on zone [$\bar{1}$010]. The (10$\bar{1}$0) direction in reciprocal space is the in-plane direction, while the (0002) direction is the out-of-plane direction. Dotted and dashed circles indicate the CoM of the center disk on the X$_{2}$O$_{3}$ and Y$_{2}$O$_{3}$ side of the interface, respectively. The black circle indicates the mask. (e) Absolute thickness mapping of the cross-sectional lamellae measured through EELS. (f) Zoom-in of absolute thickness mapping measured through EELS from dotted rectangle in (e). (g) Calculated E-field. (h) Zoom-in of E-field from dotted rectangle in (g). The scale bars in (a-c) and (e-h) indicate 50 nm.}
    \label{fig:4dstem_dpc}
\end{figure}

To experimentally probe the coupling of interfacial electrostatics and corrosion, we grew heterostructures of Fe$_{2}$O$_{3}$ and Cr$_{2}$O$_{3}$ with varying interfacial chemical microstructure, as demonstrated in prior work ~\cite{Kaspar2016}. Using oxygen-plasma-assisted molecular beam epitaxy (OPA-MBE), two types of Fe$_{2}$O$_{3}$-Cr$_{2}$O$_{3}$ heterostructures were grown by intentionally alternating the order to produce both abrupt (100 nm Fe$_{2}$O$_{3}$ / 100 nm Cr$_{2}$O$_{3}$ / Al$_{2}$O$_{3}$(0001)) and mixed (100 nm Cr$_{2}$O$_{3}$/ 100 nm Fe$_{2}$O$_{3}$ / 15 nm Cr$_{2}$O$_{3}$ / Al$_{2}$O$_{3}$(0001)) interfaces. A 15 nm Cr$_{2}$O$_{3}$ buffer layer was included below the Cr$_{2}$O$_{3}$ / Fe$_{2}$O$_{3}$ film to facilitate the deposition of Fe$_{2}$O$_{3}$ under the same lattice mismatch conditions as the Fe$_{2}$O$_{3}$ / Cr$_{2}$O$_{3}$ film. Here we take the convention that oxides are listed from top to bottom (i.e. top-oxide / middle-oxide / bottom-oxide / substrate). For each interface type, the top layer was irradiated with 100 keV Fe$^{+}$ for one sample and the other was left pristine. The irradiated samples consisted of a mixed interface with irradiated Cr$_{2}$O$_{3}$ and an abrupt interface with irradiated Fe$_{2}$O$_{3}$. High angle annular dark-field (HAADF-STEM) images of the cross-sections of the irradiated layered heterostructures are shown in Fig. \ref{fig:4dstem_dpc}(a),(b), revealing oxide film layers of approximately 100 nm. 

The four Fe$_{2}$O$_{3}$-Cr$_{2}$O$_{3}$ heterostructure samples were further characterized by the precession electron diffraction-based differential phase contrast (PED-DPC) method, a novel technique enabled by 4D-STEM. This high resolution characterization method was used to measure electrostatic effects at each interface of sample cross sections prepared by focused ion beam (FIB) liftout (Fig. \ref{fig:4dstem_dpc}(a),(b) and Supplementary Information). Using precession e-beam illumination (Fig. \ref{fig:4dstem_dpc}(c)), we measured unique diffraction patterns (averaged through beam-tilting angles) individually at each pixel over the entire scan region, which assists in suppressing dynamical effects and enhancing the signal to noise ratio. The mean diffraction pattern of the irradiated Fe$_{2}$O$_{3}$ / Cr$_{2}$O$_{3}$ abrupt interface is shown in Fig. \ref{fig:4dstem_dpc}(d). As indicated by the positional shift in center-of-mass (CoM) of the center disk shown in Fig. \ref{fig:4dstem_dpc}(d), the change in electrostatic effects between layers occurring across the interface corresponds to a direct change in interfacial electric field. By normalizing the CoM shift with the absolute thickness of each sample, measured using electron energy loss spectroscopy (EELS) (Fig. \ref{fig:4dstem_dpc}(e),(f)), the electric field across each sample was quantified based on equations (\ref{eq:e_perp}) and (\ref{eq:e_xz}) (see Methods section) as shown in Fig. \ref{fig:4dstem_dpc}(g),(h).

\begin{figure}[h]
    \centering
    \includegraphics[width=1.0\linewidth]{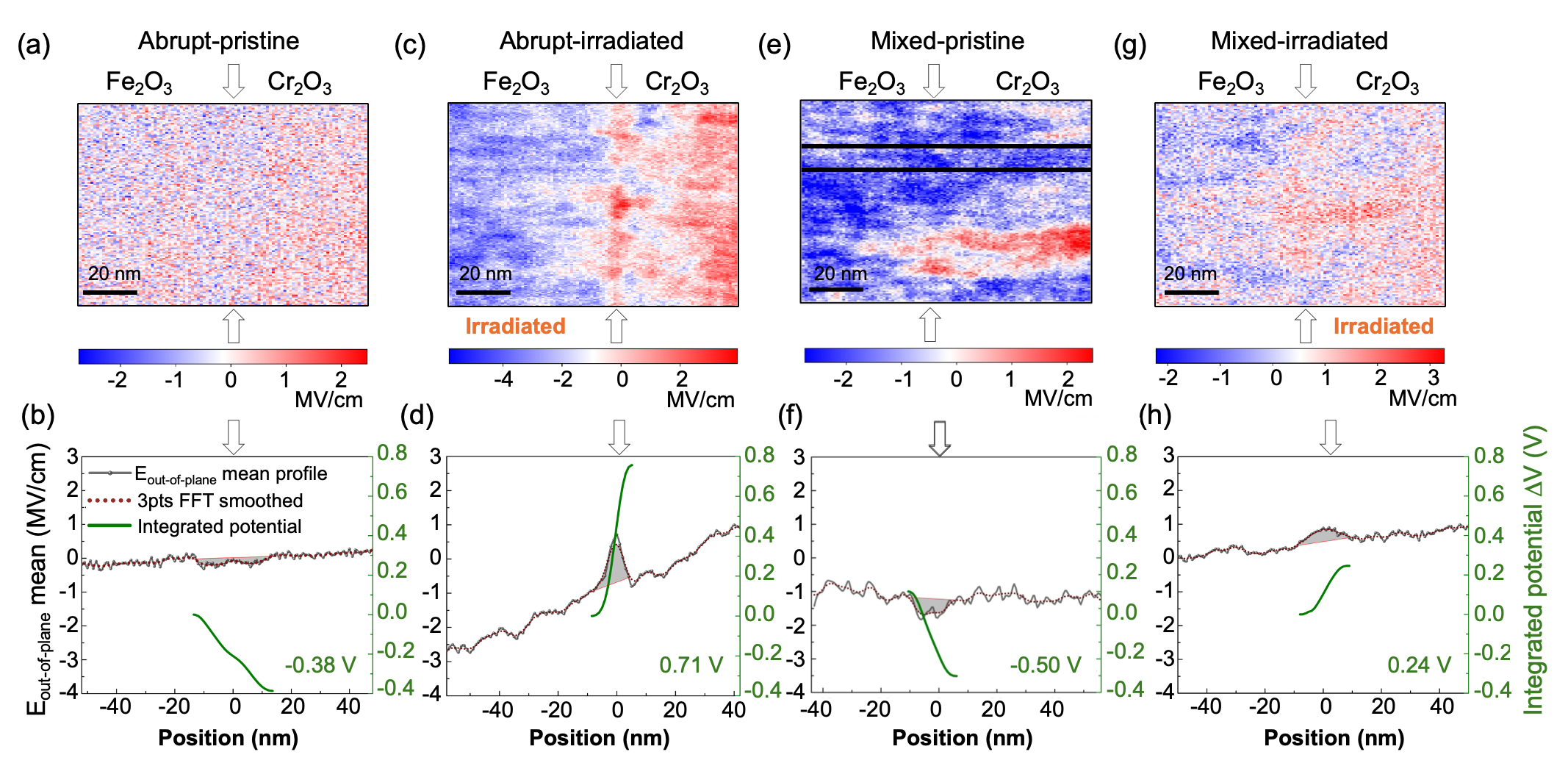}
    \caption{\textbf{The out-of-plane electric fields and integrated potentials at the oxide interfaces.} (a), (c), (e), and (g) are the electric distribution maps (MV/cm) for each of the four heterostructure samples with white arrows indicating the position of the interface. In irradiated samples, the side of the interface that was irradiated is indicated in orange text. In the mixed-pristine case, the area used to integrate the electric field to find the potential is enclosed in the black lines. (b), (d), (f), and (h) are plotted profiles of the out-of-plane mean electric field (left y-axis, gray line) and mean integrated potential (MIP) (right y-axis, blue line) plotted with respect to position along the (0001) axis (nm). The dotted line (burgundy) is a smoothed mean profile of E-field (3-points fast Fourier transform filter). The dashed line (green) denotes the change in MIP at the interface. Scale bars indicate 20 nm.}
    \label{fig:e_fields_potentials}
\end{figure}

The potential difference across each interface within 50 nm of the interface on either side was calculated by integrating the electric field obtained from the CoM shifts. The out-of-plane component of the electric field (E$_{z}$, perpendicular to the interface, aligned along the (0002) direction) was calculated from the reciprocal space center-of-mass shift, $\vec{k}_{CoM}(z)$, using equation (\ref{eq:e_perp}) (see Methods section). For each experimental condition, 2D field maps  along with the corresponding 1D profiles (averaged parallel along the interface) of the mean electric field and integrated potential are presented in Fig. \ref{fig:e_fields_potentials}. The spatial coordinate ($z$-axis) corresponds to the scanning direction across the interface, while the white arrows mark the nominal interfacial position. To quantify the potential step across each interface, the FFT-smoothed field profile was integrated across $\sim$50 nm centered at the interface.

For the pristine samples, the integrated potential across the interface was negative for both the abrupt interface (Fe$_{2}$O$_{3}$ / Cr$_{2}$O$_{3}$, Fig. \ref{fig:e_fields_potentials}(a),(b)) and the mixed interface (Cr$_{2}$O$_{3}$ / Fe$_{2}$O$_{3}$, Fig. \ref{fig:e_fields_potentials}(e),(f)) when crossing from the Fe$_{2}$O$_{3}$ side to the Cr$_{2}$O$_{3}$ side of the interface. Quantitatively, the integrated potential across the abrupt interface ($-0.38$ V) is smaller than the integrated potential across the mixed interface ($-0.50$ V). This is in qualitative agreement with the observations measured by Kaspar et al. ~\cite{Kaspar2016} via XPS; notably, for the abrupt pristine case the integrated potential is in quantitative agreement with these prior XPS measurements. 

The mixed pristine interface exhibits substantial variability in the electric field as a function of in-plane position (the y-axis in the figure). To account for this, the electric field was integrated over a smaller, more uniform, in-plane region, indicated by the black lines in Fig. \ref{fig:e_fields_potentials}(e)(f). The source of the inhomogeneity at the interface is not entirely understood. We suspect that factors including clustering of highly-localized point defects or grain boundaries in a small region near the interface may have produced an unexpectedly large jump in electric field, as seen in the strong blue to red transition in the lower half of Fig. \ref{fig:e_fields_potentials}(e). We do not expect this region to be representative of the mixed pristine Cr$_{2}$O$_{3}$ / Fe$_{2}$O$_{3}$ interface.

After irradiation the potential change is substantial, becoming positive in both samples but with a more significant change occurring for the abrupt interface (Fig. \ref{fig:e_fields_potentials}(c),(d)) compared to the mixed interface (Fig. \ref{fig:e_fields_potentials}(g),(h)). The 2D map of the Fe$_{2}$O$_{3}$ / Cr$_{2}$O$_{3}$ abrupt interface shows a well-defined bipolar structure with a strong red–blue contrast across the interface; field values reach a maxima of $+2.5$ MV/cm on the Cr$_{2}$O$_{3}$ side and minima of $−3.0$ MV/cm on the Fe$_{2}$O$_{3}$ side. The mean profile confirms a sharp transition in E$_{z}$ from negative to positive within $\sim$15 nm of the interface, suggesting strong field localization. The integrated potential difference across the interfacial region reaches $+$0.71 V, almost twice the magnitude of the pristine case but with reversed sign. This significant shift implies the formation of a substantial built-in field, likely arising from radiation-induced fixed charges, defect dipoles, or space charge layers at or near the interface. After irradiation, the Cr$_{2}$O$_{3}$ / Fe$_{2}$O$_{3}$ mixed interface (Fig. \ref{fig:e_fields_potentials}(g),(h)) shows a smaller but nontrivial increase in field contrast. The field map displays local variations up to $\pm$2 MV/cm. The mean profile shows modest fluctuations across the interface than in the abrupt case and the integrated potential reaches  $+$0.24 V.

To quantitatively analyze the interfacial charge distribution across the Fe$_{2}$O$_{3}$ / Cr$_{2}$O$_{3}$ and Cr$_{2}$O$_{3}$ / Fe$_{2}$O$_{3}$ oxide interfaces using 4D-STEM DPC analysis, we extracted the projected electric field profiles (E$_{z}$) and evaluated their spatial derivatives column-wise to obtain planar charge density using Gauss’s law:

\begin{equation}
\rho(z) = \varepsilon_{0}\frac{dE_{z}}{dz}
\end{equation}

\noindent where $\varepsilon_{0}$ is the vacuum permittivity and $\rho$ is the volume charge density in 3 dimensions, with units C/m$^{3}$. We neglect the material-specific relative permittivities, $\varepsilon_{r}$, as they are very similar, $\approx12$ for Fe$_{2}$O$_{3}$ ~\cite{Lide2007} and $\approx11.9$ for Cr$_{2}$O$_{3}$ ~\cite{Fang1963}; qualitatively, the calculated results should not be affected. After integrating along the beam direction using the projected thickness obtained via EELS, we obtained 2D sheet charge densities with units C/m$^{2}$. 

\begin{figure}
    \centering
    \includegraphics[width=1.0\linewidth]{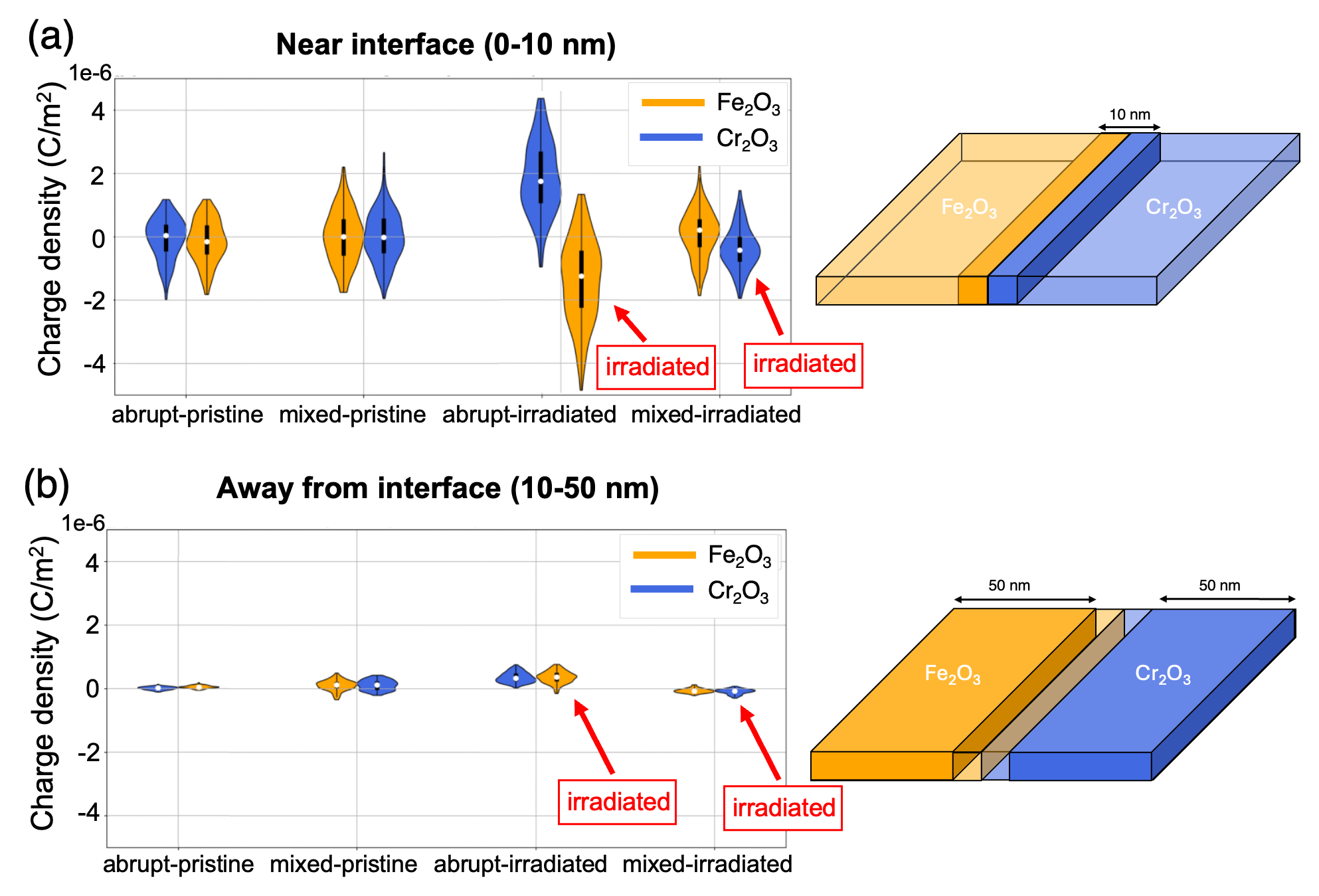}
    \caption{\textbf{The charge density along 2D sheets obtained from integrating the electric field across the oxide interfaces using Gauss's law.}(a) Plots of the statistical density of the charge density in close proximity (10 nm) to the interface for the four test cases. (b) Plots of the statistical density of charge density integrated over a broad (50 nm) region on either side of the interface. The irradiated oxide is indicated in red boxes and arrows. Charge densities are separately shown for Fe$_{2}$O$_{3}$ (gold) and Cr$_{2}$O$_{3}$ (blue) sides of the interface, as indicated in the schematics (right). In all cases, the charge densities are ordered with the bottom oxide on the left and the top oxide on the right.}
    \label{fig:charge_density_transfer}
\end{figure}

The average charge density measured with respect to sample depth (distance from the interface in the out-of-plane direction) is plotted for scan regions in close proximity (10 nm) to the interface and extending farther away (50 nm) from the interface in Fig. \ref{fig:charge_density_transfer}. The width of each violin plot represents the relative statistical density of charge density values that fall within the measured region on either the Fe$_{2}$O$_{3}$ (gold) or Cr$_{2}$O$_{3}$ (blue) side of the interface. The white dot in the middle of each violin indicates the mean charge density. This charge density consist of contributions from both electron charge and charged point defects. Within 10 nm of the interface (Fig. \ref{fig:charge_density_transfer}(a)), there is substantial charge density build-up at the interface. The pristine Fe$_{2}$O$_{3}$ / Cr$_{2}$O$_{3}$ abrupt interface has mean charge densities of $-0.22$ C/m$^{2}$ on the Fe$_{2}$O$_{3}$ side and $+0.26$ C/m$^{2}$ on the Cr$_{2}$O$_{3}$ side, with standard deviations of approximately $\pm0.04$ C/m$^{2}$, indicating measurable local charge imbalance. The pristine Cr$_{2}$O$_{3}$ / Fe$_{2}$O$_{3}$ mixed interface displays a more symmetric distribution; although there is a slight imbalance, both oxide layers have mean values closer to zero ($-0.05$ and $+0.06$ C/m$^{2}$, respectively). This is consistent with the smaller band offsets observed via XPS and DFT.

Post-irradiation, both abrupt and mixed interfaces exhibit much broader electric field profile distributions (Fig. \ref{fig:e_fields_potentials}(c)(d)(g)(h)). The irradiated Fe$_{2}$O$_{3}$ / Cr$_{2}$O$_{3}$ abrupt interface shows increased fluctuation and asymmetry (Fig. \ref{fig:e_fields_potentials}(c)); on the Fe$_{2}$O$_{3}$ side the mean charge density is $-1.23$ C/m$^{2}$ and on the Cr$_{2}$O$_{3}$ side the charge density is $+1.74$ C/m$^{2}$ (Fig. \ref{fig:charge_density_transfer}(a), third column). This indicates that negatively charged radiation-induced defects and electrons have accumulated in Fe$_{2}$O$_{3}$ (and positive charges in Cr$_{2}$O$_{3}$), inducing an increased interfacial electric field relative to the pristine sample. The irradiated Cr$_{2}$O$_{3}$ / Fe$_{2}$O$_{3}$ mixed interface shows the opposite propensity for charge density redistribution; on the Fe$_{2}$O$_{3}$ side the mean charge density is $+0.22$ C/m$^{2}$ and on the Cr$_{2}$O$_{3}$ side the mean charge density is $-0.42$ C/m$^{2}$. (Fig. \ref{fig:charge_density_transfer}(a), fourth column) The net charge density disparity is smaller. Interestingly, in both cases sources of negative charge density (e.g. electrons, cation vacancies) prefer to reside in the irradiated oxide, whereas sources of positive charge density (e.g. holes, anion vacancies) prefer to reside in the unirradiated oxide. This suggests that irradiation of the top layer induces an electric field of the same orientation irrespective of the ordering of the oxides; the magnitude of that electric field and charge redistribution is dependent on the interfacial atomistic chemical structure. 

When integrated over a broader $\pm50$ nm window (Fig. \ref{fig:charge_density_transfer}(b)), all regions show a convergence toward net neutrality. Average values range from $+0.04$ to $-0.06$ C/m$^{2}$, and the standard deviations decrease accordingly. This suggests that the net interface dipole is localized, and long-range electrostatic balance is preserved.

\begin{figure}
    \centering
    \includegraphics[width=0.7\linewidth]{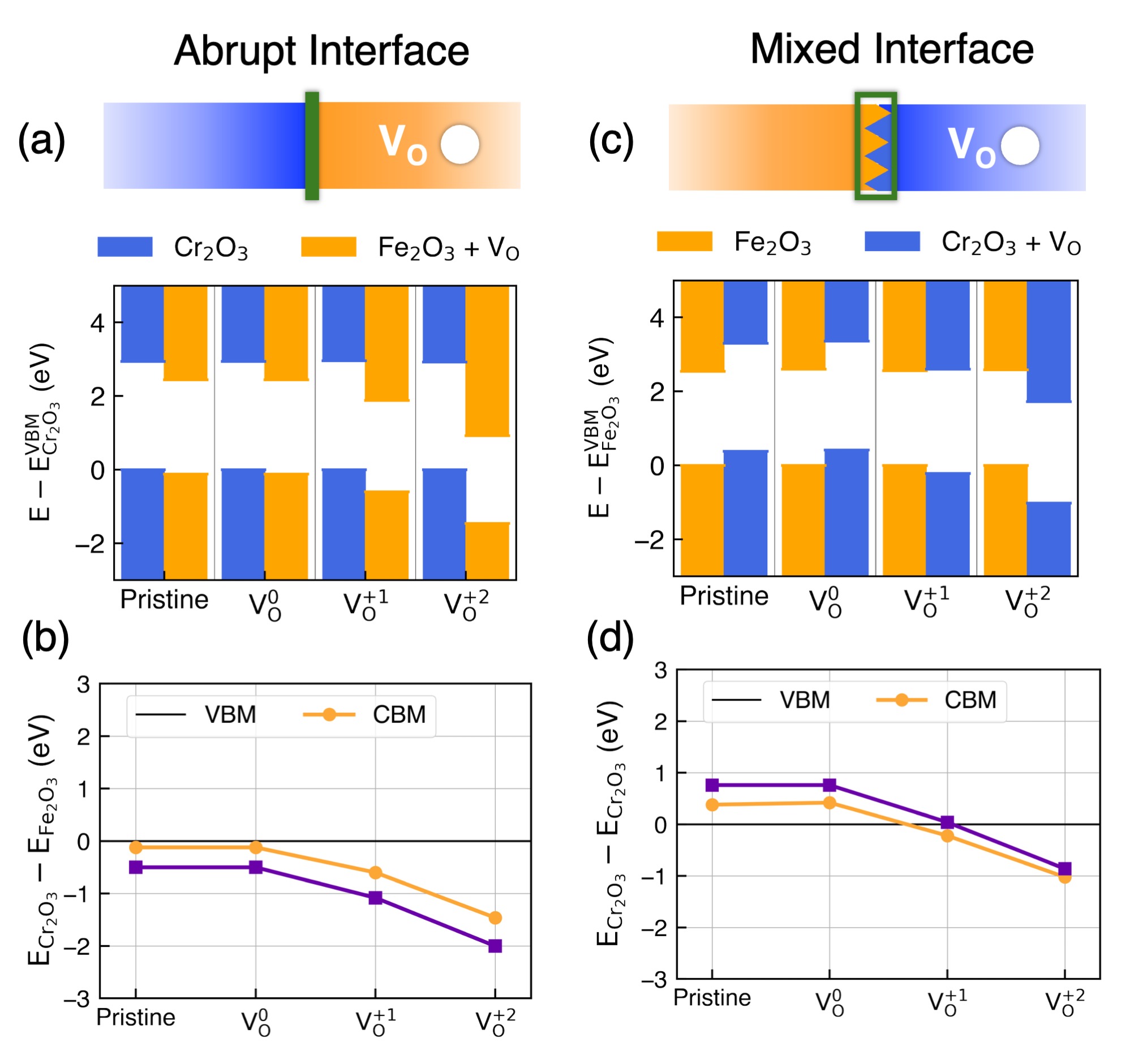}
    \caption{\textbf{The relative band offsets (neglecting in-gap defect states) of the abrupt and mixed interface heterostructures with oxygen vacancies 1.4 nm from the interface.} For (a)(b) the abrupt interface, the Fe$_{2}$O$_{3}$ layer contains the oxygen vacancy. The Fe$_{2}$O$_{3}$ band edges shift down relative to the band edges of Cr$_{2}$O$_{3}$, with increasing magnitude as the oxygen vacancy charge state increases. For (c)(d) the mixed interface, the Cr$_{2}$O$_{3}$ layer contains the oxygen vacancy. The Cr$_{2}$O$_{3}$ band edges shift down relative to the band edges of Fe$_{2}$O$_{3}$, also with increasing magnitude as the oxygen vacancy charge state increases.}
    \label{fig:band_offsets}
\end{figure}

Our experimental measurements indicate that the interfacial electric fields for interfaces with different atomistic structures respond to irradiation differently. To gain more insight into how interfacial atomistic structure couples with irradiation-induced defects, we performed DFT calculations of Fe$_{2}$O$_{3}$-Cr$_{2}$O$_{3}$ heterostructures with oxygen vacancy point defects, as one representative defect that would form under irradiation. This choice was also motivated by the fact that oxygen defects tend to migrate faster than cation defects and are more likely to reach, or cross, the interface ~\cite{Yano2021b, Banerjee2023}. Oxygen vacancy (V$_{\mathrm{O}}$) point defects were introduced in each oxygen layer of Fe$_{2}$O$_{3}$-Cr$_{2}$O$_{3}$ heterostructures with 6 cation layers each ($2\times2\times1$ supercells of each oxide with 12 cation layers total). Both neutral and positively charged oxygen vacancies were considered (V$_{\mathrm{O}}^{0}$, V$_{\mathrm{O}}^{+1}$, V$_{\mathrm{O}}^{+2}$). We calculated the density of states under conditions analogous to the experimentally measured irradiated heterostructures: an abrupt interface with oxygen vacancies in the Fe$_{2}$O$_{3}$ layer and a mixed interface with oxygen vacancies in the Cr$_{2}$O$_{3}$ layer. Oxygen vacancies produce highly localized in-gap states. The highly local nature of these in-gap states allowed us to again use the layer resolved density of states to extract the band edge positions of the bulk regions of Fe$_{2}$O$_{3}$ and Cr$_{2}$O$_{3}$ for comparison to the bulk band edges of the pristine structures (see Supplementary Information Fig. S5 and S6).

 Oxygen vacancies were placed 1.4 nm away from the interface, a distance close enough to represent defects clustered near the interface, but far enough away that the atomistic structure (abrupt vs. mixed) of the interface was unchanged. Charged point defects induced a dramatic change in the relative band offsets at the Fe$_{2}$O$_{3}$ - Cr$_{2}$O$_{3}$ interface. For the abrupt interface, where the experimental sample had an irradiated Fe$_{2}$O$_{3}$ top layer, the Fe$_{2}$O$_{3}$ bands shifted downward relative to the Cr$_{2}$O$_{3}$ band edges (Fig. \ref{fig:band_offsets}(a)(b)) Similarly, for the mixed interface, where the experimental sample had an irradiated Cr$_{2}$O$_{3}$ top layer, the Cr$_{2}$O$_{3}$ bands shifted downward relative to the Fe$_{2}$O$_{3}$ band edges (Fig. \ref{fig:band_offsets}(c)(d)). In both cases these band edge shifts favor negative (electron) charge carriers moving from the oxide with no defects to the oxide with oxygen vacancies. The distinguishing feature between the two atomistic structures is the magnitude of the final band offset, which is larger for the abrupt interface than the mixed interface. These results are qualitatively consistent with the experimentally observed directions of charge segregation and relative magnitudes of charge segregation at the interfaces of the irradiated samples; it is favorable for negative charge density to shift to the irradiated oxide.

 To confirm that defects do not cluster so closely to the interface that the atomistic structure of the interface is destroyed, we calculated the band offsets when oxygen vacancies were placed immediately at the interface (0.14 nm from the interfacial cation layer). Both the abrupt and mixed interfaces showed minimal difference in the bulk band offset across the interface (see Supplementary Information, Fig. S7). Coupled with the experimental charge segregation data, this strongly indicates that the atomistic chemical details of the interfaces (abrupt vs. mixed) are maintained, not destroyed, by the introduction of irradiation-induced defects.

We expect many types of defects, such as  cation vacancies and interstitials, to be present in the irradiated samples in addition to oxygen vacancies, which are not accounted for in these calculations. However, even an examination of one type of defect reinforces the experimentally measured impact of defects on the electronic structure and band offsets of this class of oxide heterointerface.

\section{\label{sec:discussion}Discussion}

Our high-resolution 4D-STEM DPC characterization and DFT simulations demonstrate that irradiation-induced defects and the atomistic interfacial chemical structure of Fe$_{2}$O$_{3}$ - Cr$_{2}$O$_{3}$ heterostructures couple to produce substantial modulation of interfacial electric fields, with implications for the thermokinetics and spatial distribution of defects and resulting corrosion mechanisms.

The dependence of band offset in the pristine structures on the chemical details of the interfacial structure that we calculate via DFT is consistent with prior XPS experimental results that demonstrated variation of built-in potentials. Correspondingly, through 4D-STEM DPC analysis, we uncover evidence of a built-in electric field and potential at the pristine Fe$_{2}$O$_{3}$ / Cr$_{2}$O$_{3}$ and Cr$_{2}$O$_{3}$ / Fe$_{2}$O$_{3}$ interfaces, with the magnitude of the built-in electric field dependent on the atomistic chemical structure of the interface. 

Our DFT calculations of the effect of charged oxygen vacancy defects on band edge offset and our experimental 4D-STEM DPC measurements of charge density transfer across the Fe$_{2}$O$_{3}$-Cr$_{2}$O$_{3}$ interface show remarkable qualitative consistency. Both predict that irradiation amplifies the built-in electric field and points it in the same direction out-of-plane, irrespective of whether Fe$_{2}$O$_{3}$ or Cr$_{2}$O$_{3}$ was irradiated; our results reveal that irradiation by Fe$^{+}$ ions promotes the population of negative charges (e.g. electrons, cation vacancies) in the irradiated layer and promotes a population of positive charges (e.g. holes, anion vacancies) in the unirradiated layer, amplifying the interfacial electric field. The magnitude of this amplification is dependent on the chemical details of the interfacial structure. Whether the direction of the charge segregation is impacted by the charge of the particles used for irradiation (in this case positive) is a question we leave for future work.

Using band offset as a proxy for the electrostatic thermodynamic barrier for charge carriers to migrate from one oxide to the other, our DFT results show that the migration of defects, most significantly charged defects, should be appreciably affected by the interfacial atomistic structure, consistent with the experimental 4D-STEM DPC results. Our DFT calculations also indicate that point defects introduced by radiation damage in the bulk have a greater effect on the electric field than those that cluster at the interface. This indicates that maintaining a population of point defects in the bulk should preserve the propensity and longevity of irradiation-induced interfacial electric fields.  

Practical application of these findings would require identifying which type of charged defect is the most prolific (i.e., whether there is a higher population of positive or negative charged defects) in a heterostructure of interest to identify the dominant direction of charge transfer and interfacial electric field amplification as a function of interfacial structure. 

Our findings offer a plausible pathway to design oxide heterostructures that selectively promote populations of specific types of point defects to reside in a pre-selected oxide of choice; this has the potential to be used to produce protective coatings that leverage interfacial electrostatics to control corrosion. These results further demonstrate that understanding the underlying electronic structure at oxide interfaces is critical to understanding the dynamical coupling of irradiation and corrosion of oxide heterostructures in extreme environments.

\section{\label{sec:methods}Methods}

\subsection{\label{sec:dft_methods}First-principles calculations}

First-principles calculations were performed on heterostructures of Fe$_{2}$O$_{3}$ and Cr$_{2}$O$_{3}$ using density functional theory (DFT) with a plane-wave basis and projector augmented wave (PAW) potentials ~\cite{Kresse1999} as implemented by the Vienna $\textit{ab initio}$ Simulation Package (VASP) ~\cite{Kresse1996a, Kresse1996b}. All calculations were spin-polarized. Two supercell sizes were considered: a heterostructure of $2\times2\times1$ supercells with 6 layers each of Fe$_{2}$O$_{3}$ and Cr$_{2}$O$_{3}$ and a heterostructure of $2\times2\times2$ supercells with 12 layers each of Fe$_{2}$O$_{3}$ and Cr$_{2}$O$_{3}$.  The total energies of eight antiferromagnetic (AFM) and ferromagnetic (FM) orderings were calculated. The lowest energy magnetic configuration was identified to have FM in-plane ordering with alternating AFM between layers in Fe$_{2}$O$_{3}$ and AFM ordering both in-plane and between layers for Cr$_{2}$O$_{3}$. Structural relaxations were performed in the generalized gradient approximation as implemented by Perdew, Burke, and Ernzerhof (PBE) ~\cite{Perdew1996} with additional Hubbard U corrections (PBE + U) as implemented by Dudarev ~\cite{Dudarev1998} in order to account for charge localization on the \textit{d}-orbitals of the Fe and Cr ions. Effective U values of 5.3 eV and 3.7 eV were used for Fe and Cr, respectively. An energy cutoff of 500 eV with $12\times12\times1$ and $8\times8\times1$ $\Gamma$-centered k-grids were used for the $2\times2\times1$ and $2\times2\times2$ supercells, respectively, with total energy converged to 10$^{–5}$ eV and forces converged to $<$1 meV/\textup{\AA}. 

We first performed crystal structure relaxations on $\alpha$-Fe$_{2}$O$_{3}$ and Cr$_{2}$O$_{3}$ individually. The relaxed lattice parameters of $\alpha$-Fe$_{2}$O$_{3}$ (a = 5.084 \textup{\AA} and c = 13.877 \textup{\AA}) were in good agreement with experimental values (a = 5.032 \textup{\AA} and c = 13.764 \textup{\AA} ~\cite{Gokhan2019}. Similarly, the relaxed lattice parameters of Cr$_{2}$O$_{3}$ (a = 5.018 \textup{\AA} and c = 13.764 \textup{\AA}) were in good agreement with experimental values (a = 4.959 \textup{\AA} and c = 13.595 \textup{\AA}) ~\cite{Fabrykiewicz2018}. This corresponds to 1.3\% lattice mismatch in the DFT relaxed in-plane lattice parameter, 1.5\% in the experimental values. The abrupt and mixed heterostructures were then constructed and were each allowed to relax to a coherent compromise lattice parameter. The abrupt and mixed heterostructures had in-plane lattice parameters of a = 5.080 \textup{\AA} and a = 5.081 \textup{\AA} respectively, close to the lattice parameter of $\alpha$-Fe$_{2}$O$_{3}$.

Oxygen vacancies were introduced into each layer of the $2\times2\times1$ supercell and geometric relaxations were performed, with fixed lattice parameters, for neutral, $+$1, and $+$2 charge states. Oxygen vacancies were also introduced into the center of each oxide layer of the $2\times2\times2$ supercells and geometric relaxations were performed, with fixed lattice parameters, for neutral, $+$1, and $+$2 charge states. Density of states (DOS) calculations were performed for both the $2\times2\times1$ and $2\times2\times2$ supercells and the projections onto each cation layer were extracted to calculate the relative band edge offsets of Fe$_{2}$O$_{3}$ and Cr$_{2}$O$_{3}$.

\subsection{\label{sec:growth_methods}Epitaxial Thin-Film Growth}

Epitaxial thin films of $\alpha$-Fe$_{2}$O$_{3}$ and Cr$_{2}$O$_{3}$ were deposited on $\alpha$-Al$_{2}$O$_{3}$(0001) substrates by oxygen-plasma-assisted molecular beam epitaxy (OPA-MBE). Substrate surfaces were precleaned in situ by exposure to activated oxygen from an electron cyclotron resonance (ECR) microwave plasma source at a background oxygen pressure of 2.7 $\times$ 10$^{-3}$ Pa and room temperature for 30 min. The flow of activated oxygen continued as the substrate was heated to the deposition temperature of 700–750 $^\circ$C. High purity iron and chromium were evaporated from electron beam evaporators and each flux was monitored by atomic absorption spectroscopy.  Two epitaxial film stack geometries were synthesized: 100 nm Fe$_{2}$O$_{3}$ / 100 nm Cr$_{2}$O$_{3}$ / Al$_{2}$O$_{3}$(0001) and 100 nm Cr$_{2}$O$_{3}$ / 100 nm Fe$_{2}$O$_{3}$ / 15 nm Cr$_{2}$O$_{3}$ / Al$_{2}$O$_{3}$(0001) (all thicknesses are nominal values). We take the convention that oxides are listed from top to bottom (i.e. top-oxide / middle-oxide / bottom-oxide / substrate).  A 15 nm buffer layer of Cr$_{2}$O$_{3}$ was included to avoid the significant misfit dislocation density and associated island growth mode that arises when depositing Fe$_{2}$O$_{3}$ directly on Al$_{2}$O$_{3}$(0001) ~\cite{Chambers2000}.  The lattice mismatch between Fe$_{2}$O$_{3}$(0001) and Al$_{2}$O$_{3}$(0001) is (a$_{film}$ – a$_{substrate}$)/a$_{substrate}$ × 100\% = 5.8\%, whereas the lattice mismatch between Fe$_{2}$O$_{3}$(0001) and Cr$_{2}$O$_{3}$(0001) is only 1.5\%.  Prior studies have shown that Cr$_{2}$O$_{3}$ relaxes to near its bulk lattice parameter when deposited at thicknesses greater than 10 nm on Al$_{2}$O$_{3}$(0001) ~\cite{Chambers2000, Kaspar2023}.

\subsection{\label{sec:irradiation_methods}Thin Film Irradiation}

One film stack of each geometry was irradiated with 100 keV Fe$^{+}$ across the entire surface via a Danfysik implanter at room temperature. Stopping and Range of Ions in Matter (SRIM) simulations ~\cite{Ziegler2013} were performed uniformly using the Kinchin-Pease quick calculation mode ~\cite{Yuan2019} (displacement energies of 40 eV for Fe and Cr, 28 eV for O) and indicated that the fluence required for a damage level of 0.1 displacements per atom (dpa), averaged over the top 100 nm of the sample (the nominal thickness of the top film layer), was 7.38 $\times$ 10$^{13}$ ions cm$^{-2}$ for Fe$_{2}$O$_{3}$, and the same fluence was used for irradiation of Cr$_{2}$O$_{3}$. The objective was to ensure that the interface was not directly damaged; rather, we sought to probe whether radiation-induced defects would migrate to and interact with the interfaces after being created. As shown in the Supplemental Information (Fig. S1), the irradiation dose peaks at $\sim$0.14 dpa at a depth of 20 nm from the film surface and drops to nearly zero at the interface between the two films (nominal 100 nm depth).  Likewise, the Fe$^{+}$ implantation concentration peaks at 0.016\% at $\sim$46 nm from the film surface and drops to nearly zero at a depth of 100 nm.  From these calculations, the assumption is made that little or no dose or implantation directly occurs at the interface or in the underlying film.

\subsection{\label{sec:4dstem_dpc_methods}4D-STEM DPC Analysis}

Cross-sectional TEM lamellae for 4D-STEM DPC analysis were prepared by a conventional lift-out method using a Thermo Fisher Helios G4 dual-beam focused ion beam/scanning electron microscope (FIB/SEM). The lift-out procedure was adapted from atom probe tomography (APT) specimen preparation techniques to mitigate charge effects and prevent cut-through damage. Initially, a 0.5 $\mu$m electron-beam Pt layer was deposited to reduce charging and near-surface FIB damage, followed by a 2 $\mu$m ion-beam-deposited Pt protective cap using a 30 kV Ga$^{+}$ ion beam. The surrounding trenches were milled at 30$^\circ$ tilt on both sides using high-current FIB settings, and lamellas were extracted using an EasyLift nanomanipulator. FIB thinning was performed at 30 kV to a sample dimension of $\sim$500 nm, followed by fine thinning steps at 16 kV, 5 kV, and 2 kV to reach a final lamella thickness suitable for STEM imaging.

4D-STEM DPC datasets were acquired using a Thermo Fisher Scientific ThemIS TEM at an accelerating voltage of 300 kV. The electron probe was set with a 20 mm C3 aperture using a nominal convergence semi-angle of 0.82 mrad and a spot size of 3. Precession electron diffraction (PED) in conjunction with 4D-STEM was implemented by tilting and rotating the electron beam conically around the optical axis. A Nanomegas Stingray detector/camera was employed to record diffraction patterns at each scan position with frame rate 0.1 fps. and a precession angle of 0.499$^\circ$. This allowed for enhanced sampling in reciprocal space while reducing dynamical scattering effects. The beam spot size in real space was $\sim$1.2 nm with precession on, and the probe current was maintained at 0.04 nA. The scan area was selected to encompass the Fe$_{2}$O$_{3}$ / Cr$_{2}$O$_{3}$ interface, with a real-space step size of 1 nm that was sufficient to resolve interfacial features. Each diffraction pattern was recorded with $580\times580$ pixels resolution, and beam alignment and precession calibration were performed prior to acquisition to minimize beam tilt artifacts through Nanomegas software.

The CoM shift for each diffraction pattern was computed to quantify the average beam deflection induced by the local electric field ~\cite{Beyer2021, Toyama2022, Chejarla2023}. CoM shifts along the $x$ and $z$ directions, denoted as CoM$_{x}$ and CoM$_{z}$, respectively, were calculated after appropriate centering corrections. Here, the $x$-axis is parallel to the interface and the $z$-axis is perpendicular to the interface, i.e. along the (0001) axis. A circular mask centered on the bright field disk was applied during CoM extraction to improve the signal-to-noise ratio by excluding contributions from diffracted disks.
The CoM shift for each diffraction pattern was calculated to quantify the local beam deflection induced by the projected electric field. The CoM along the $x$ and $z$ directions was computed using the intensity-weighted average of diffraction coordinates:

\begin{equation}
\mathrm{CoM}_{\nu}=\frac{\mathrm{\sum}_{i,j} I(i.j)\cdot k_{\nu}(i,j)}{\mathrm{\sum}_{i,j}I_{i,j}}
\end{equation}

\noindent where $I(i,j)$ is the intensity at pixel $(i,j)$, $\nu$ is the cartesian direction ($x$ or $z$) and ($k_{x}(i,j)$, $k_{z}(i,j)$) are the reciprocal space coordinates of each pixel relative to the beam center. An annular mask centered on the bright-field disk was applied during integration to minimize contributions from diffracted disks and high-angle scattering. The electric field from the CoM shift can be written as 
\begin{equation}
\vec{E}_{\perp}=\frac{h}{e} \cdot \frac{1}{t(x,z)} \cdot \vec{k}_{\mathrm{CoM}}(x,z).
\label{eq:e_perp}
\end{equation}

\noindent where $t(x,z)$ is the sample thickness at each position and $\vec{k}_{\mathrm{CoM}}(x,z)$ is the measured shift in the CoM as a result of the internal electric field. The local electric field components can be converted into real-space components E$_{x}$, E$_{z}$ by
\begin{equation}
    \mathrm{E}_{x,z}= \alpha \cdot \frac{\mathrm{CoM}_{x,z}}{t(x,z)}
    \label{eq:e_xz}
\end{equation}

\noindent where $\alpha$ is a proportionality constant dependent on the electron beam energy that converts the measured beam deflection (CoM shift) into a physical electric field value by accounting for the momentum and velocity of 300 kV electrons:

\begin{equation}
    \alpha = \frac{h\nu}{e} \thickspace \mathrm{and} \thickspace \nu = \frac{1}{\lambda} = \frac{\sqrt{2m_{e}eV}}{h} .
\end{equation}

For 300 keV electrons, $\alpha$ is 489.74; this value accounts for the relativistic momentum and detector geometry of the 300 kV STEM system and is consistent with previously reported calibration standards ~\cite{Beyer2021}. 

EELS spectra were acquired with a Gatan Quantum ER system at 300 kV, using a 5 mm entrance aperture and a dispersion setting of 0.25 eV/channel. A convergence semi-angle of 10 mrad with 10 eV energy filter slit was applied. The zero-loss peak and low-loss region were acquired in dual-EELS mode. The inelastic mean free path ($\lambda$) was estimated based on composition, and the measured relative thickness values yielded mean lamellae thicknesses.  

 A constant thickness was assumed across the scanned area, based on independent thickness measurements via the electron energy loss spectroscopy (EELS) log-ratio method. The relative thickness ($\frac{t}{\lambda}$) at each scanned position was determined using:

\begin{equation}
    \frac{t}{\lambda}= \mathrm{ln} \left( \frac{I_{total}}{I_{0}} \right) 
\end{equation}

\noindent where $I_{total}$ is the total integrated EELS intensity (zero-loss + plasmon) and $I_{0}$ is the zero-loss peak intensity. To convert relative thickness into absolute thickness (in nm), an effective inelastic mean free path ($\lambda$) was estimated based on the sample composition and experimental conditions. The value was obtained from an established model, the modified Malis formula, with literature values for Cr$_{2}$O$_{3}$ and Fe$_{2}$O$_{3}$ ~\cite{Powell2002,Egerton2011}, accounting for density and band gap corrections. The log ratio of the $I_{total}$/$I_{0}$ was measured to be one, which led us to use the calculated MFP to be the thickness of the sample. The inelastic mean free path in the Malis formula is written as

\begin{equation}
    \lambda=\frac{106\times E_{0}}{\rho \sqrt{Z}} \left( 1+\frac{E_{g}}{E_{0}} \right)^{-1}
\end{equation}

\noindent where $E_{0}$ is the beam energy (keV), $\rho$ is the density (g/cm$^{3}$),  $Z$ is the effective atomic number, and  $E_{g}$ is the bandgap (eV). The calculated thicknesses for the four samples analyzed were as follows: abrupt-pristine 287nm; abrupt-irradiated 254 nm;  mixed-pristine 274 nm ;  mixed-irradiated 236nm. This calculated absolute thickness $t$ was then used in the conversion of CoM shifts to electric fields.

In this analysis, the position of the Fe$_{2}$O$_{3}$ / Cr$_{2}$O$_{3}$ interface was determined by correlating the bright-field intensity contrast and the discontinuity in CoM shift maps. A binary mask was generated to separate the regions above (Cr$_{2}$O$_{3}$ side) and below (Fe$_{2}$O$_{3}$ side) the interface (or vice versa) for subsequent analysis. Column-wise electric field profiles were extracted along the beam propagation direction across the interface. For each column, two separate linear fits were performed: one for the region above the interface and one for the region below.

\section{\label{sec:ackn}Acknowledgements}
E.A.P., H.K., Y.W., and B.P.U. acknowledge support to perform this research from Los Alamos National Laboratory, which is operated by Triad National Security, LLC, for the National Nuclear Security Administration of the U.S. Department of Energy (Contract No. 89233218CNA000001). T.C.K. acknowledges support to perform this research from Pacific Northwest National Laboratory, a multiprogram national laboratory operated by Battelle for the U.S. DOE under Contract DE-AC05-79RL01830. H.K. and Y.W. acknowledge support to perform ion beam irradiation at the Center for Integrated Nanotechnologies (CINT), a DOE Office of Science User Facility. E.A.P. acknowledges computational resources supported in part by CINT, a DOE Office of Science user facility, in partnership with the LANL Institutional Computing Program for computational resources. Additional calculations were performed at the National Energy Research Scientific Computing Center (NERSC), a U.S. Department of Energy Office of Science User Facility located at Lawrence Berkeley National Laboratory, operated under Contract No. DE-AC02-05CH11231 using NERSC award ERCAP0031336. D.L., S.H.M. and A.M.M. received support from the Molecular Foundry, a user facility operated by the Office of Science, Office of Basic Energy Sciences, of the U.S. Department of Energy under Contract No. DE-AC02-05CH11231.

\section{\label{sec:ackn}Funding sources}
E.A.P., D.L., S.H.M., T.C.K., H.K., Y.W., B.P.U., and A.M.M. disclose support for research of this work from the U.S. Department of Energy, Office of Science, Basic Energy Sciences (BES) via the Energy Frontier Research Center (EFRC) FUTURE (Fundamental Understanding of Transport Under Reactor Extremes). 

\section{\label{sec:ackn}Author contributions}
First-principles DFT calculations were planned and performed by E.A.P and B.P.U. Epitaxial film growth was performed by T.C.K. Ion beam irradiation was performed by H.K. and Y.W. 4D-STEM and DPC measurements were performed by D.L. and S.H.M. under the supervision of A.M.M.

\section{\label{sec:ackn}Competing interests}
The authors declare no competing interests. 


\begin{thebibliography}{52}%
\makeatletter
\providecommand \@ifxundefined [1]{%
 \@ifx{#1\undefined}
}%
\providecommand \@ifnum [1]{%
 \ifnum #1\expandafter \@firstoftwo
 \else \expandafter \@secondoftwo
 \fi
}%
\providecommand \@ifx [1]{%
 \ifx #1\expandafter \@firstoftwo
 \else \expandafter \@secondoftwo
 \fi
}%
\providecommand \natexlab [1]{#1}%
\providecommand \enquote  [1]{``#1''}%
\providecommand \bibnamefont  [1]{#1}%
\providecommand \bibfnamefont [1]{#1}%
\providecommand \citenamefont [1]{#1}%
\providecommand \href@noop [0]{\@secondoftwo}%
\providecommand \href [0]{\begingroup \@sanitize@url \@href}%
\providecommand \@href[1]{\@@startlink{#1}\@@href}%
\providecommand \@@href[1]{\endgroup#1\@@endlink}%
\providecommand \@sanitize@url [0]{\catcode `\\12\catcode `\$12\catcode `\&12\catcode `\#12\catcode `\^12\catcode `\_12\catcode `\%12\relax}%
\providecommand \@@startlink[1]{}%
\providecommand \@@endlink[0]{}%
\providecommand \url  [0]{\begingroup\@sanitize@url \@url }%
\providecommand \@url [1]{\endgroup\@href {#1}{\urlprefix }}%
\providecommand \urlprefix  [0]{URL }%
\providecommand \Eprint [0]{\href }%
\providecommand \doibase [0]{https://doi.org/}%
\providecommand \selectlanguage [0]{\@gobble}%
\providecommand \bibinfo  [0]{\@secondoftwo}%
\providecommand \bibfield  [0]{\@secondoftwo}%
\providecommand \translation [1]{[#1]}%
\providecommand \BibitemOpen [0]{}%
\providecommand \bibitemStop [0]{}%
\providecommand \bibitemNoStop [0]{.\EOS\space}%
\providecommand \EOS [0]{\spacefactor3000\relax}%
\providecommand \BibitemShut  [1]{\csname bibitem#1\endcsname}%
\let\auto@bib@innerbib\@empty
\bibitem [{\citenamefont {Chambers}\ \emph {et~al.}(2000)\citenamefont {Chambers}, \citenamefont {Liang},\ and\ \citenamefont {Gao}}]{Chambers2000}%
  \BibitemOpen
  \bibfield  {author} {\bibinfo {author} {\bibfnamefont {S.~A.}\ \bibnamefont {Chambers}}, \bibinfo {author} {\bibfnamefont {Y.}~\bibnamefont {Liang}},\ and\ \bibinfo {author} {\bibfnamefont {Y.}~\bibnamefont {Gao}},\ }\bibinfo {title} {Noncummutative band offset at $\alpha$-Cr$_{2}$O$_{3}$/$\alpha$-Fe$_{2}$O$_{3}$ (0001) heterojunctions},\ \href {https://doi.org/10.1103/PhysRevB.61.13223} {\bibfield  {journal} {\bibinfo  {journal} {Physical Review B}\ }\textbf {\bibinfo {volume} {61}},\ \bibinfo {pages} {13223} (\bibinfo {year} {2000})}\BibitemShut {NoStop}%
\bibitem [{\citenamefont {Jaffe}\ \emph {et~al.}(2004)\citenamefont {Jaffe}, \citenamefont {Dupuis},\ and\ \citenamefont {Gutowski}}]{Jaffe2004}%
  \BibitemOpen
  \bibfield  {author} {\bibinfo {author} {\bibfnamefont {J.~E.}\ \bibnamefont {Jaffe}}, \bibinfo {author} {\bibfnamefont {M.}~\bibnamefont {Dupuis}},\ and\ \bibinfo {author} {\bibfnamefont {M.}~\bibnamefont {Gutowski}},\ }\bibinfo {title} {First-principles study of noncommutative band offsets at $\alpha$Cr$_{2}$O$_{3}$/$\alpha$Fe$_{2}$O$_{3}$(0001) interfaces},\ \href {https://doi.org/10.1103/PhysRevB.69.205106} {\bibfield  {journal} {\bibinfo  {journal} {Physical Review B}\ }\textbf {\bibinfo {volume} {69}},\ \bibinfo {pages} {205106} (\bibinfo {year} {2004})}\BibitemShut {NoStop}%
\bibitem [{\citenamefont {Kaspar}\ \emph {et~al.}(2016)\citenamefont {Kaspar}, \citenamefont {Schreiber}, \citenamefont {Spurgeon}, \citenamefont {McBriarty}, \citenamefont {Carroll}, \citenamefont {Gamelin},\ and\ \citenamefont {Chambers}}]{Kaspar2016}%
  \BibitemOpen
  \bibfield  {author} {\bibinfo {author} {\bibfnamefont {T.~C.}\ \bibnamefont {Kaspar}}, \bibinfo {author} {\bibfnamefont {D.~K.}\ \bibnamefont {Schreiber}}, \bibinfo {author} {\bibfnamefont {S.~R.}\ \bibnamefont {Spurgeon}}, \bibinfo {author} {\bibfnamefont {M.~E.}\ \bibnamefont {McBriarty}}, \bibinfo {author} {\bibfnamefont {G.~M.}\ \bibnamefont {Carroll}}, \bibinfo {author} {\bibfnamefont {D.~R.}\ \bibnamefont {Gamelin}},\ and\ \bibinfo {author} {\bibfnamefont {S.~A.}\ \bibnamefont {Chambers}},\ }\bibinfo {title} {Built-In Potential in Fe$_{2}$O$_{3}$-Cr$_{2}$O$_{3}$ Superlattices for Improved Photoexcited Carrier Separation},\ \href {https://doi.org/https://doi.org/10.1002/adma.201504545} {\bibfield  {journal} {\bibinfo  {journal} {Advanced Materials}\ }\textbf {\bibinfo {volume} {28}},\ \bibinfo {pages} {1616} (\bibinfo {year} {2016})},\ \Eprint {https://arxiv.org/abs/https://advanced.onlinelibrary.wiley.com/doi/pdf/10.1002/adma.201504545}
  {https://advanced.onlinelibrary.wiley.com/doi/pdf/10.1002/adma.201504545} \BibitemShut {NoStop}%
\bibitem [{\citenamefont {Song}\ \emph {et~al.}(2022)\citenamefont {Song}, \citenamefont {Shi}, \citenamefont {Lu}, \citenamefont {Wang}, \citenamefont {Hu}, \citenamefont {Gao}, \citenamefont {Luo},\ and\ \citenamefont {Ma}}]{Song2022}%
  \BibitemOpen
  \bibfield  {author} {\bibinfo {author} {\bibfnamefont {A.}~\bibnamefont {Song}}, \bibinfo {author} {\bibfnamefont {R.}~\bibnamefont {Shi}}, \bibinfo {author} {\bibfnamefont {H.}~\bibnamefont {Lu}}, \bibinfo {author} {\bibfnamefont {X.}~\bibnamefont {Wang}}, \bibinfo {author} {\bibfnamefont {Y.}~\bibnamefont {Hu}}, \bibinfo {author} {\bibfnamefont {H.-J.}\ \bibnamefont {Gao}}, \bibinfo {author} {\bibfnamefont {J.}~\bibnamefont {Luo}},\ and\ \bibinfo {author} {\bibfnamefont {T.}~\bibnamefont {Ma}},\ }\bibinfo {title} {Fluctuation of Interfacial Electronic Properties Induces Friction Tuning under an Electric Field},\ \href {https://doi.org/10.1021/acs.nanolett.1c04116} {\bibfield  {journal} {\bibinfo  {journal} {Nano Letters}\ }\textbf {\bibinfo {volume} {22}},\ \bibinfo {pages} {1889} (\bibinfo {year} {2022})},\ \bibinfo {note} {pMID: 35171620},\ \Eprint {https://arxiv.org/abs/https://doi.org/10.1021/acs.nanolett.1c04116} {https://doi.org/10.1021/acs.nanolett.1c04116} \BibitemShut {NoStop}%
\bibitem [{\citenamefont {Ophus}(2023)}]{Ophus2023}%
  \BibitemOpen
  \bibfield  {author} {\bibinfo {author} {\bibfnamefont {C.}~\bibnamefont {Ophus}},\ }\bibinfo {title} {Quantitative Scanning Transmission Electron Microscopy for Materials Science: Imaging, Diffraction, Spectroscopy, and Tomography},\ \href {https://doi.org/https://doi.org/10.1146/annurev-matsci-080921-092646} {\bibfield  {journal} {\bibinfo  {journal} {Annual Review of Materials Research}\ }\textbf {\bibinfo {volume} {53}},\ \bibinfo {pages} {105} (\bibinfo {year} {2023})}\BibitemShut {NoStop}%
\bibitem [{\citenamefont {Li}\ \emph {et~al.}(2024)\citenamefont {Li}, \citenamefont {Mu}, \citenamefont {Korytov}, \citenamefont {Alexandrou},\ and\ \citenamefont {Bosch}}]{Chen2024}%
  \BibitemOpen
  \bibfield  {author} {\bibinfo {author} {\bibfnamefont {C.}~\bibnamefont {Li}}, \bibinfo {author} {\bibfnamefont {X.}~\bibnamefont {Mu}}, \bibinfo {author} {\bibfnamefont {M.}~\bibnamefont {Korytov}}, \bibinfo {author} {\bibfnamefont {I.}~\bibnamefont {Alexandrou}},\ and\ \bibinfo {author} {\bibfnamefont {E.~G.~T.}\ \bibnamefont {Bosch}},\ }\bibinfo {title} {Differential phase contrast (DPC) mapping electric fields: Optimising experimental conditions},\ \href {https://doi.org/https://doi.org/10.1111/jmi.13271} {\bibfield  {journal} {\bibinfo  {journal} {Journal of Microscopy}\ }\textbf {\bibinfo {volume} {293}},\ \bibinfo {pages} {177} (\bibinfo {year} {2024})},\ \Eprint {https://arxiv.org/abs/https://onlinelibrary.wiley.com/doi/pdf/10.1111/jmi.13271} {https://onlinelibrary.wiley.com/doi/pdf/10.1111/jmi.13271} \BibitemShut {NoStop}%
\bibitem [{\citenamefont {da~Silva}\ \emph {et~al.}(2022)\citenamefont {da~Silva}, \citenamefont {Momtaz}, \citenamefont {Bruas}, \citenamefont {Rouvi\'{e}re}, \citenamefont {Okuno}, \citenamefont {Cooper},\ and\ \citenamefont {den Hertog}}]{daSilva2022}%
  \BibitemOpen
  \bibfield  {author} {\bibinfo {author} {\bibfnamefont {B.~C.}\ \bibnamefont {da~Silva}}, \bibinfo {author} {\bibfnamefont {Z.~S.}\ \bibnamefont {Momtaz}}, \bibinfo {author} {\bibfnamefont {L.}~\bibnamefont {Bruas}}, \bibinfo {author} {\bibfnamefont {J.-L.}\ \bibnamefont {Rouvi\'{e}re}}, \bibinfo {author} {\bibfnamefont {H.}~\bibnamefont {Okuno}}, \bibinfo {author} {\bibfnamefont {D.}~\bibnamefont {Cooper}},\ and\ \bibinfo {author} {\bibfnamefont {M.~I.}\ \bibnamefont {den Hertog}},\ }\bibinfo {title} {The influence of illumination conditions in the measurement of built-in electric field at p–n junctions by 4D-STEM},\ \href {https://doi.org/10.1063/5.0104861} {\bibfield  {journal} {\bibinfo  {journal} {Applied Physics Letters}\ }\textbf {\bibinfo {volume} {121}},\ \bibinfo {pages} {123503} (\bibinfo {year} {2022})}\BibitemShut {NoStop}%
\bibitem [{\citenamefont {Smith}\ \emph {et~al.}(2025)\citenamefont {Smith}, \citenamefont {Tran}, \citenamefont {Roccapriore}, \citenamefont {Shen}, \citenamefont {Zhang},\ and\ \citenamefont {Chi}}]{Smith2025}%
  \BibitemOpen
  \bibfield  {author} {\bibinfo {author} {\bibfnamefont {J.}~\bibnamefont {Smith}}, \bibinfo {author} {\bibfnamefont {H.}~\bibnamefont {Tran}}, \bibinfo {author} {\bibfnamefont {K.~M.}\ \bibnamefont {Roccapriore}}, \bibinfo {author} {\bibfnamefont {Z.}~\bibnamefont {Shen}}, \bibinfo {author} {\bibfnamefont {G.}~\bibnamefont {Zhang}},\ and\ \bibinfo {author} {\bibfnamefont {M.}~\bibnamefont {Chi}},\ }\bibinfo {title} {Advanced Compressive Sensing and Dynamic Sampling for 4D-STEM Imaging of Interfaces},\ \href {https://doi.org/https://doi.org/10.1002/smtd.202400742} {\bibfield  {journal} {\bibinfo  {journal} {Small Methods}\ }\textbf {\bibinfo {volume} {9}},\ \bibinfo {pages} {2400742} (\bibinfo {year} {2025})},\ \Eprint {https://arxiv.org/abs/https://onlinelibrary.wiley.com/doi/pdf/10.1002/smtd.202400742} {https://onlinelibrary.wiley.com/doi/pdf/10.1002/smtd.202400742} \BibitemShut {NoStop}%
\bibitem [{\citenamefont {Lee}\ \emph {et~al.}(2022)\citenamefont {Lee}, \citenamefont {Jeon},\ and\ \citenamefont {Lee}}]{Lee2022}%
  \BibitemOpen
  \bibfield  {author} {\bibinfo {author} {\bibfnamefont {S.}~\bibnamefont {Lee}}, \bibinfo {author} {\bibfnamefont {J.}~\bibnamefont {Jeon}},\ and\ \bibinfo {author} {\bibfnamefont {H.}~\bibnamefont {Lee}},\ }\bibinfo {title} {Probing oxygen vacancy distribution in oxide heterostructures by deep Learning-based spectral analysis of current noise},\ \href {https://doi.org/https://doi.org/10.1016/j.apsusc.2022.154599} {\bibfield  {journal} {\bibinfo  {journal} {Applied Surface Science}\ }\textbf {\bibinfo {volume} {604}},\ \bibinfo {pages} {154599} (\bibinfo {year} {2022})}\BibitemShut {NoStop}%
\bibitem [{\citenamefont {Kruska}\ \emph {et~al.}(2012)\citenamefont {Kruska}, \citenamefont {Lozano-Perez}, \citenamefont {Saxey}, \citenamefont {Terachi}, \citenamefont {Yamada},\ and\ \citenamefont {Smith}}]{Kruska2012}%
  \BibitemOpen
  \bibfield  {author} {\bibinfo {author} {\bibfnamefont {K.}~\bibnamefont {Kruska}}, \bibinfo {author} {\bibfnamefont {S.}~\bibnamefont {Lozano-Perez}}, \bibinfo {author} {\bibfnamefont {D.~W.}\ \bibnamefont {Saxey}}, \bibinfo {author} {\bibfnamefont {T.}~\bibnamefont {Terachi}}, \bibinfo {author} {\bibfnamefont {T.}~\bibnamefont {Yamada}},\ and\ \bibinfo {author} {\bibfnamefont {G.~D.}\ \bibnamefont {Smith}},\ }\bibinfo {title} {Nanoscale characterisation of grain boundary oxidation in cold-worked stainless steels},\ \href {https://doi.org/https://doi.org/10.1016/j.corsci.2012.06.030} {\bibfield  {journal} {\bibinfo  {journal} {Corrosion Science}\ }\textbf {\bibinfo {volume} {63}},\ \bibinfo {pages} {225} (\bibinfo {year} {2012})}\BibitemShut {NoStop}%
\bibitem [{\citenamefont {Samin}\ \emph {et~al.}(2019{\natexlab{a}})\citenamefont {Samin}, \citenamefont {Andersson}, \citenamefont {Holby},\ and\ \citenamefont {Uberuaga}}]{Samin2019a}%
  \BibitemOpen
  \bibfield  {author} {\bibinfo {author} {\bibfnamefont {A.~J.}\ \bibnamefont {Samin}}, \bibinfo {author} {\bibfnamefont {D.~A.}\ \bibnamefont {Andersson}}, \bibinfo {author} {\bibfnamefont {E.~F.}\ \bibnamefont {Holby}},\ and\ \bibinfo {author} {\bibfnamefont {B.~P.}\ \bibnamefont {Uberuaga}},\ }\bibinfo {title} {First-principles localized cluster expansion study of the kinetics of hydrogen diffusion in homogeneous and heterogeneous Fe-Cr alloys},\ \href {https://doi.org/10.1103/PhysRevB.99.014110} {\bibfield  {journal} {\bibinfo  {journal} {Physical Review B}\ }\textbf {\bibinfo {volume} {99}},\ \bibinfo {pages} {014110} (\bibinfo {year} {2019}{\natexlab{a}})}\BibitemShut {NoStop}%
\bibitem [{\citenamefont {Samin}\ \emph {et~al.}(2019{\natexlab{b}})\citenamefont {Samin}, \citenamefont {Andersson}, \citenamefont {Holby},\ and\ \citenamefont {Uberuaga}}]{Samin2019b}%
  \BibitemOpen
  \bibfield  {author} {\bibinfo {author} {\bibfnamefont {A.~J.}\ \bibnamefont {Samin}}, \bibinfo {author} {\bibfnamefont {D.~A.}\ \bibnamefont {Andersson}}, \bibinfo {author} {\bibfnamefont {E.~F.}\ \bibnamefont {Holby}},\ and\ \bibinfo {author} {\bibfnamefont {B.~P.}\ \bibnamefont {Uberuaga}},\ }\bibinfo {title} {Ab initio based examination of the kinetics and thermodynamics of oxygen in Fe-Cr alloys},\ \href {https://doi.org/10.1103/PhysRevB.99.174202} {\bibfield  {journal} {\bibinfo  {journal} {Physical Review B}\ }\textbf {\bibinfo {volume} {99}},\ \bibinfo {pages} {174202} (\bibinfo {year} {2019}{\natexlab{b}})}\BibitemShut {NoStop}%
\bibitem [{\citenamefont {Schmidt}\ \emph {et~al.}(2021)\citenamefont {Schmidt}, \citenamefont {Hosemann}, \citenamefont {Scarlat}, \citenamefont {Schreiber}, \citenamefont {Scully},\ and\ \citenamefont {Uberuaga}}]{Schmidt2021}%
  \BibitemOpen
  \bibfield  {author} {\bibinfo {author} {\bibfnamefont {F.}~\bibnamefont {Schmidt}}, \bibinfo {author} {\bibfnamefont {P.}~\bibnamefont {Hosemann}}, \bibinfo {author} {\bibfnamefont {R.~O.}\ \bibnamefont {Scarlat}}, \bibinfo {author} {\bibfnamefont {D.~K.}\ \bibnamefont {Schreiber}}, \bibinfo {author} {\bibfnamefont {J.~R.}\ \bibnamefont {Scully}},\ and\ \bibinfo {author} {\bibfnamefont {B.~P.}\ \bibnamefont {Uberuaga}},\ }\bibinfo {title} {Effects of Radiation-Induced Defects on Corrosion},\ \href {https://doi.org/10.1146/annurev-matsci-080819-123403} {\bibfield  {journal} {\bibinfo  {journal} {Annual Review of Materials Research}\ }\textbf {\bibinfo {volume} {51}},\ \bibinfo {pages} {293} (\bibinfo {year} {2021})}\BibitemShut {NoStop}%
\bibitem [{\citenamefont {Haseman}\ \emph {et~al.}(2021)\citenamefont {Haseman}, \citenamefont {Somodi}, \citenamefont {Stepanov}, \citenamefont {Wall}, \citenamefont {Boatner}, \citenamefont {Hosemann}, \citenamefont {Wang}, \citenamefont {Uberuaga},\ and\ \citenamefont {Selim}}]{Haseman2021}%
  \BibitemOpen
  \bibfield  {author} {\bibinfo {author} {\bibfnamefont {M.}~\bibnamefont {Haseman}}, \bibinfo {author} {\bibfnamefont {C.~B.}\ \bibnamefont {Somodi}}, \bibinfo {author} {\bibfnamefont {P.}~\bibnamefont {Stepanov}}, \bibinfo {author} {\bibfnamefont {D.~E.}\ \bibnamefont {Wall}}, \bibinfo {author} {\bibfnamefont {L.~A.}\ \bibnamefont {Boatner}}, \bibinfo {author} {\bibfnamefont {P.}~\bibnamefont {Hosemann}}, \bibinfo {author} {\bibfnamefont {Y.~Q.}\ \bibnamefont {Wang}}, \bibinfo {author} {\bibfnamefont {B.~P.}\ \bibnamefont {Uberuaga}},\ and\ \bibinfo {author} {\bibfnamefont {F.~A.}\ \bibnamefont {Selim}},\ }\bibinfo {title} {Neutron irradiation induced defects in oxides and their impact on the oxide properties},\ \href {https://doi.org/10.1063/5.0046292} {\bibfield  {journal} {\bibinfo  {journal} {Journal of Applied Physics}\ }\textbf {\bibinfo {volume} {129}},\ \bibinfo {pages} {215901} (\bibinfo {year} {2021})}\BibitemShut {NoStop}%
\bibitem [{\citenamefont {Lach}\ \emph {et~al.}(2021)\citenamefont {Lach}, \citenamefont {Olszta}, \citenamefont {Taylor}, \citenamefont {Yano}, \citenamefont {Edwards}, \citenamefont {Byun}, \citenamefont {Chou},\ and\ \citenamefont {Schreiber}}]{Lach2021}%
  \BibitemOpen
  \bibfield  {author} {\bibinfo {author} {\bibfnamefont {T.~G.}\ \bibnamefont {Lach}}, \bibinfo {author} {\bibfnamefont {M.~J.}\ \bibnamefont {Olszta}}, \bibinfo {author} {\bibfnamefont {S.~D.}\ \bibnamefont {Taylor}}, \bibinfo {author} {\bibfnamefont {K.~H.}\ \bibnamefont {Yano}}, \bibinfo {author} {\bibfnamefont {D.~J.}\ \bibnamefont {Edwards}}, \bibinfo {author} {\bibfnamefont {T.~S.}\ \bibnamefont {Byun}}, \bibinfo {author} {\bibfnamefont {P.~H.}\ \bibnamefont {Chou}},\ and\ \bibinfo {author} {\bibfnamefont {D.~K.}\ \bibnamefont {Schreiber}},\ }\bibinfo {title} {Correlative STEM-APT characterization of radiation-induced segregation and precipitation of in-service BWR 304 stainless steel},\ \href {https://doi.org/https://doi.org/10.1016/j.jnucmat.2021.152894} {\bibfield  {journal} {\bibinfo  {journal} {Journal of Nuclear Materials}\ }\textbf {\bibinfo {volume} {549}},\ \bibinfo {pages} {152894} (\bibinfo {year} {2021})}\BibitemShut {NoStop}%
\bibitem [{\citenamefont {Agarwal}\ \emph {et~al.}(2022)\citenamefont {Agarwal}, \citenamefont {Butterling}, \citenamefont {Liedke}, \citenamefont {Yano}, \citenamefont {Schreiber}, \citenamefont {Jones}, \citenamefont {Uberuaga}, \citenamefont {Wang}, \citenamefont {Chancey}, \citenamefont {Kim}, \citenamefont {Derby}, \citenamefont {Li}, \citenamefont {Edwards}, \citenamefont {Hosemann}, \citenamefont {Kaoumi}, \citenamefont {Hirschmann}, \citenamefont {Wagner},\ and\ \citenamefont {Selim}}]{Agarwal2022}%
  \BibitemOpen
  \bibfield  {author} {\bibinfo {author} {\bibfnamefont {S.}~\bibnamefont {Agarwal}}, \bibinfo {author} {\bibfnamefont {M.}~\bibnamefont {Butterling}}, \bibinfo {author} {\bibfnamefont {M.~O.}\ \bibnamefont {Liedke}}, \bibinfo {author} {\bibfnamefont {K.~H.}\ \bibnamefont {Yano}}, \bibinfo {author} {\bibfnamefont {D.~K.}\ \bibnamefont {Schreiber}}, \bibinfo {author} {\bibfnamefont {A.~C.~L.}\ \bibnamefont {Jones}}, \bibinfo {author} {\bibfnamefont {B.~P.}\ \bibnamefont {Uberuaga}}, \bibinfo {author} {\bibfnamefont {Y.~Q.}\ \bibnamefont {Wang}}, \bibinfo {author} {\bibfnamefont {M.}~\bibnamefont {Chancey}}, \bibinfo {author} {\bibfnamefont {H.}~\bibnamefont {Kim}}, \bibinfo {author} {\bibfnamefont {B.~K.}\ \bibnamefont {Derby}}, \bibinfo {author} {\bibfnamefont {N.}~\bibnamefont {Li}}, \bibinfo {author} {\bibfnamefont {D.~J.}\ \bibnamefont {Edwards}}, \bibinfo {author} {\bibfnamefont {P.}~\bibnamefont {Hosemann}}, \bibinfo {author} {\bibfnamefont {D.}~\bibnamefont {Kaoumi}}, \bibinfo {author} {\bibfnamefont
  {E.}~\bibnamefont {Hirschmann}}, \bibinfo {author} {\bibfnamefont {A.}~\bibnamefont {Wagner}},\ and\ \bibinfo {author} {\bibfnamefont {F.~A.}\ \bibnamefont {Selim}},\ }\bibinfo {title} {The mechanism behind the high radiation tolerance of Fe–Cr alloys},\ \href {https://doi.org/10.1063/5.0085086} {\bibfield  {journal} {\bibinfo  {journal} {Journal of Applied Physics}\ }\textbf {\bibinfo {volume} {131}},\ \bibinfo {pages} {125903} (\bibinfo {year} {2022})}\BibitemShut {NoStop}%
\bibitem [{\citenamefont {Agarwal}\ \emph {et~al.}(2020)\citenamefont {Agarwal}, \citenamefont {Liedke}, \citenamefont {Jones}, \citenamefont {Reed}, \citenamefont {Kohnert}, \citenamefont {Uberuaga}, \citenamefont {Wang}, \citenamefont {Cooper}, \citenamefont {Kaoumi}, \citenamefont {Li}, \citenamefont {Auguste}, \citenamefont {Hosemann}, \citenamefont {Capolungo}, \citenamefont {Edwards}, \citenamefont {Butterling}, \citenamefont {Hirschmann}, \citenamefont {Wagner},\ and\ \citenamefont {Selim}}]{Agarwal2020}%
  \BibitemOpen
  \bibfield  {author} {\bibinfo {author} {\bibfnamefont {S.}~\bibnamefont {Agarwal}}, \bibinfo {author} {\bibfnamefont {M.~O.}\ \bibnamefont {Liedke}}, \bibinfo {author} {\bibfnamefont {A.~C.~L.}\ \bibnamefont {Jones}}, \bibinfo {author} {\bibfnamefont {E.}~\bibnamefont {Reed}}, \bibinfo {author} {\bibfnamefont {A.~A.}\ \bibnamefont {Kohnert}}, \bibinfo {author} {\bibfnamefont {B.~P.}\ \bibnamefont {Uberuaga}}, \bibinfo {author} {\bibfnamefont {Y.~Q.}\ \bibnamefont {Wang}}, \bibinfo {author} {\bibfnamefont {J.}~\bibnamefont {Cooper}}, \bibinfo {author} {\bibfnamefont {D.}~\bibnamefont {Kaoumi}}, \bibinfo {author} {\bibfnamefont {N.}~\bibnamefont {Li}}, \bibinfo {author} {\bibfnamefont {R.}~\bibnamefont {Auguste}}, \bibinfo {author} {\bibfnamefont {P.}~\bibnamefont {Hosemann}}, \bibinfo {author} {\bibfnamefont {L.}~\bibnamefont {Capolungo}}, \bibinfo {author} {\bibfnamefont {D.~J.}\ \bibnamefont {Edwards}}, \bibinfo {author} {\bibfnamefont {M.}~\bibnamefont {Butterling}}, \bibinfo {author} {\bibfnamefont
  {E.}~\bibnamefont {Hirschmann}}, \bibinfo {author} {\bibfnamefont {A.}~\bibnamefont {Wagner}},\ and\ \bibinfo {author} {\bibfnamefont {F.~A.}\ \bibnamefont {Selim}},\ }\bibinfo {title} {A new mechanism for void-cascade interaction from nondestructive depth-resolved atomic-scale measurements of ion irradiation–induced defects in Fe},\ \href {https://doi.org/10.1126/sciadv.aba8437} {\bibfield  {journal} {\bibinfo  {journal} {Science Advances}\ }\textbf {\bibinfo {volume} {6}},\ \bibinfo {pages} {eaba8437} (\bibinfo {year} {2020})},\ \Eprint {https://arxiv.org/abs/https://www.science.org/doi/pdf/10.1126/sciadv.aba8437} {https://www.science.org/doi/pdf/10.1126/sciadv.aba8437} \BibitemShut {NoStop}%
\bibitem [{\citenamefont {Yano}\ \emph {et~al.}(2021{\natexlab{a}})\citenamefont {Yano}, \citenamefont {Kohnert}, \citenamefont {Kaspar}, \citenamefont {Taylor}, \citenamefont {Spurgeon}, \citenamefont {Kim}, \citenamefont {Wang}, \citenamefont {Uberuaga},\ and\ \citenamefont {Schreiber}}]{Yano2021a}%
  \BibitemOpen
  \bibfield  {author} {\bibinfo {author} {\bibfnamefont {K.~H.}\ \bibnamefont {Yano}}, \bibinfo {author} {\bibfnamefont {A.~A.}\ \bibnamefont {Kohnert}}, \bibinfo {author} {\bibfnamefont {T.~C.}\ \bibnamefont {Kaspar}}, \bibinfo {author} {\bibfnamefont {S.~D.}\ \bibnamefont {Taylor}}, \bibinfo {author} {\bibfnamefont {S.~R.}\ \bibnamefont {Spurgeon}}, \bibinfo {author} {\bibfnamefont {H.}~\bibnamefont {Kim}}, \bibinfo {author} {\bibfnamefont {Y.}~\bibnamefont {Wang}}, \bibinfo {author} {\bibfnamefont {B.~P.}\ \bibnamefont {Uberuaga}},\ and\ \bibinfo {author} {\bibfnamefont {D.~K.}\ \bibnamefont {Schreiber}},\ }\bibinfo {title} {Radiation Enhanced Anion Diffusion in Chromia},\ \href {https://doi.org/10.1021/acs.jpcc.1c08705} {\bibfield  {journal} {\bibinfo  {journal} {The Journal of Physical Chemistry C}\ }\textbf {\bibinfo {volume} {125}},\ \bibinfo {pages} {27820} (\bibinfo {year} {2021}{\natexlab{a}})},\ \Eprint {https://arxiv.org/abs/https://doi.org/10.1021/acs.jpcc.1c08705}
  {https://doi.org/10.1021/acs.jpcc.1c08705} \BibitemShut {NoStop}%
\bibitem [{\citenamefont {Yano}\ \emph {et~al.}(2021{\natexlab{b}})\citenamefont {Yano}, \citenamefont {Kohnert}, \citenamefont {Banerjee}, \citenamefont {Edwards}, \citenamefont {Holby}, \citenamefont {Kaspar}, \citenamefont {Kim}, \citenamefont {Lach}, \citenamefont {Taylor}, \citenamefont {Wang}, \citenamefont {Uberuaga},\ and\ \citenamefont {Schreiber}}]{Yano2021b}%
  \BibitemOpen
  \bibfield  {author} {\bibinfo {author} {\bibfnamefont {K.~H.}\ \bibnamefont {Yano}}, \bibinfo {author} {\bibfnamefont {A.~A.}\ \bibnamefont {Kohnert}}, \bibinfo {author} {\bibfnamefont {A.}~\bibnamefont {Banerjee}}, \bibinfo {author} {\bibfnamefont {D.~J.}\ \bibnamefont {Edwards}}, \bibinfo {author} {\bibfnamefont {E.~F.}\ \bibnamefont {Holby}}, \bibinfo {author} {\bibfnamefont {T.~C.}\ \bibnamefont {Kaspar}}, \bibinfo {author} {\bibfnamefont {H.}~\bibnamefont {Kim}}, \bibinfo {author} {\bibfnamefont {T.~G.}\ \bibnamefont {Lach}}, \bibinfo {author} {\bibfnamefont {S.~D.}\ \bibnamefont {Taylor}}, \bibinfo {author} {\bibfnamefont {Y.~Q.}\ \bibnamefont {Wang}}, \bibinfo {author} {\bibfnamefont {B.~P.}\ \bibnamefont {Uberuaga}},\ and\ \bibinfo {author} {\bibfnamefont {D.~K.}\ \bibnamefont {Schreiber}},\ }\bibinfo {title} {Radiation-Enhanced Anion Transport in Hematite},\ \href {https://doi.org/10.1021/acs.chemmater.0c04235} {\bibfield  {journal} {\bibinfo  {journal} {Chemistry of Materials}\ }\textbf {\bibinfo
  {volume} {33}},\ \bibinfo {pages} {2307} (\bibinfo {year} {2021}{\natexlab{b}})},\ \Eprint {https://arxiv.org/abs/https://doi.org/10.1021/acs.chemmater.0c04235} {https://doi.org/10.1021/acs.chemmater.0c04235} \BibitemShut {NoStop}%
\bibitem [{\citenamefont {Yano}\ \emph {et~al.}(2022)\citenamefont {Yano}, \citenamefont {Kohnert}, \citenamefont {Kaspar}, \citenamefont {Taylor}, \citenamefont {Spurgeon}, \citenamefont {Kim}, \citenamefont {Wang}, \citenamefont {Uberuaga},\ and\ \citenamefont {Schreiber}}]{Yano2022}%
  \BibitemOpen
  \bibfield  {author} {\bibinfo {author} {\bibfnamefont {K.~H.}\ \bibnamefont {Yano}}, \bibinfo {author} {\bibfnamefont {A.~A.}\ \bibnamefont {Kohnert}}, \bibinfo {author} {\bibfnamefont {T.~C.}\ \bibnamefont {Kaspar}}, \bibinfo {author} {\bibfnamefont {S.~D.}\ \bibnamefont {Taylor}}, \bibinfo {author} {\bibfnamefont {S.~R.}\ \bibnamefont {Spurgeon}}, \bibinfo {author} {\bibfnamefont {H.}~\bibnamefont {Kim}}, \bibinfo {author} {\bibfnamefont {Y.~Q.}\ \bibnamefont {Wang}}, \bibinfo {author} {\bibfnamefont {B.~P.}\ \bibnamefont {Uberuaga}},\ and\ \bibinfo {author} {\bibfnamefont {D.~K.}\ \bibnamefont {Schreiber}},\ }\bibinfo {title} {Dose rate dependent cation \& anion radiation enhanced diffusion in hematite},\ \href {https://doi.org/10.1039/D2TA03403D} {\bibfield  {journal} {\bibinfo  {journal} {Journal Materials Chemistry A}\ }\textbf {\bibinfo {volume} {10}},\ \bibinfo {pages} {24167} (\bibinfo {year} {2022})}\BibitemShut {NoStop}%
\bibitem [{\citenamefont {Owusu-Mensah}\ \emph {et~al.}(2022)\citenamefont {Owusu-Mensah}, \citenamefont {Cooper}, \citenamefont {Lopez~Morales}, \citenamefont {Yano}, \citenamefont {Taylor}, \citenamefont {Schreiber}, \citenamefont {Uberuaga},\ and\ \citenamefont {Kaoumi}}]{OwusuMensah2022}%
  \BibitemOpen
  \bibfield  {author} {\bibinfo {author} {\bibfnamefont {M.}~\bibnamefont {Owusu-Mensah}}, \bibinfo {author} {\bibfnamefont {J.}~\bibnamefont {Cooper}}, \bibinfo {author} {\bibfnamefont {A.}~\bibnamefont {Lopez~Morales}}, \bibinfo {author} {\bibfnamefont {K.}~\bibnamefont {Yano}}, \bibinfo {author} {\bibfnamefont {S.~D.}\ \bibnamefont {Taylor}}, \bibinfo {author} {\bibfnamefont {D.~K.}\ \bibnamefont {Schreiber}}, \bibinfo {author} {\bibfnamefont {B.~P.}\ \bibnamefont {Uberuaga}},\ and\ \bibinfo {author} {\bibfnamefont {D.}~\bibnamefont {Kaoumi}},\ }\bibinfo {title} {Surprisingly high irradiation-induced defect mobility in Fe$_{3}$O$_{4}$ as revealed through in situ transmission electron microscopy},\ \href {https://doi.org/https://doi.org/10.1016/j.matchar.2022.111863} {\bibfield  {journal} {\bibinfo  {journal} {Materials Characterization}\ }\textbf {\bibinfo {volume} {187}},\ \bibinfo {pages} {111863} (\bibinfo {year} {2022})}\BibitemShut {NoStop}%
\bibitem [{\citenamefont {Chan}\ \emph {et~al.}(2023)\citenamefont {Chan}, \citenamefont {Auguste}, \citenamefont {Romanovskaia}, \citenamefont {Lopez~Morales}, \citenamefont {Schmidt}, \citenamefont {Romanovski}, \citenamefont {Winkler}, \citenamefont {Qiu}, \citenamefont {Wang}, \citenamefont {Kaoumi}, \citenamefont {Selim}, \citenamefont {Uberuaga}, \citenamefont {Hosemann},\ and\ \citenamefont {Scully}}]{Chan2023}%
  \BibitemOpen
  \bibfield  {author} {\bibinfo {author} {\bibfnamefont {H.~L.}\ \bibnamefont {Chan}}, \bibinfo {author} {\bibfnamefont {R.}~\bibnamefont {Auguste}}, \bibinfo {author} {\bibfnamefont {E.}~\bibnamefont {Romanovskaia}}, \bibinfo {author} {\bibfnamefont {A.}~\bibnamefont {Lopez~Morales}}, \bibinfo {author} {\bibfnamefont {F.}~\bibnamefont {Schmidt}}, \bibinfo {author} {\bibfnamefont {V.}~\bibnamefont {Romanovski}}, \bibinfo {author} {\bibfnamefont {C.}~\bibnamefont {Winkler}}, \bibinfo {author} {\bibfnamefont {J.}~\bibnamefont {Qiu}}, \bibinfo {author} {\bibfnamefont {Y.~Q.}\ \bibnamefont {Wang}}, \bibinfo {author} {\bibfnamefont {D.}~\bibnamefont {Kaoumi}}, \bibinfo {author} {\bibfnamefont {F.~A.}\ \bibnamefont {Selim}}, \bibinfo {author} {\bibfnamefont {B.~P.}\ \bibnamefont {Uberuaga}}, \bibinfo {author} {\bibfnamefont {P.}~\bibnamefont {Hosemann}},\ and\ \bibinfo {author} {\bibfnamefont {J.~R.}\ \bibnamefont {Scully}},\ }\bibinfo {title} {Multi–length scale characterization of point defects in thermally
  oxidized, proton irradiated iron oxides},\ \href {https://doi.org/https://doi.org/10.1016/j.mtla.2023.101762} {\bibfield  {journal} {\bibinfo  {journal} {Materialia}\ }\textbf {\bibinfo {volume} {28}},\ \bibinfo {pages} {101762} (\bibinfo {year} {2023})}\BibitemShut {NoStop}%
\bibitem [{\citenamefont {Xi}\ \emph {et~al.}(2024)\citenamefont {Xi}, \citenamefont {Zhang}, \citenamefont {Su}, \citenamefont {Wei}, \citenamefont {Hu}, \citenamefont {Queylat}, \citenamefont {Kim}, \citenamefont {Couet},\ and\ \citenamefont {Szlufarska}}]{Xi2024}%
  \BibitemOpen
  \bibfield  {author} {\bibinfo {author} {\bibfnamefont {J.}~\bibnamefont {Xi}}, \bibinfo {author} {\bibfnamefont {H.}~\bibnamefont {Zhang}}, \bibinfo {author} {\bibfnamefont {R.}~\bibnamefont {Su}}, \bibinfo {author} {\bibfnamefont {S.}~\bibnamefont {Wei}}, \bibinfo {author} {\bibfnamefont {X.}~\bibnamefont {Hu}}, \bibinfo {author} {\bibfnamefont {B.}~\bibnamefont {Queylat}}, \bibinfo {author} {\bibfnamefont {T.}~\bibnamefont {Kim}}, \bibinfo {author} {\bibfnamefont {A.}~\bibnamefont {Couet}},\ and\ \bibinfo {author} {\bibfnamefont {I.}~\bibnamefont {Szlufarska}},\ }\bibinfo {title} {Coupling of radiation and grain boundary corrosion in SiC},\ \bibfield  {journal} {\bibinfo  {journal} {npj Materials Degradation}\ }\textbf {\bibinfo {volume} {8}},\ \href {https://doi.org/10.1038/s41529-024-00436-y} {10.1038/s41529-024-00436-y} (\bibinfo {year} {2024})\BibitemShut {NoStop}%
\bibitem [{\citenamefont {Lopez~Morales}\ \emph {et~al.}(2024)\citenamefont {Lopez~Morales}, \citenamefont {Chen},\ and\ \citenamefont {Kaoumi}}]{LopezMorales2024}%
  \BibitemOpen
  \bibfield  {author} {\bibinfo {author} {\bibfnamefont {A.~M.}\ \bibnamefont {Lopez~Morales}}, \bibinfo {author} {\bibfnamefont {W.-Y.}\ \bibnamefont {Chen}},\ and\ \bibinfo {author} {\bibfnamefont {D.}~\bibnamefont {Kaoumi}},\ }\bibinfo {title} {The self-annealing of irradiation induced defects in magnetite Fe$_{3}$O$_{4}$: Revealing reversible irradiation-induced disorder transformation through in situ TEM},\ \href {https://doi.org/10.1063/5.0226606} {\bibfield  {journal} {\bibinfo  {journal} {Journal of Applied Physics}\ }\textbf {\bibinfo {volume} {136}},\ \bibinfo {pages} {075903} (\bibinfo {year} {2024})}\BibitemShut {NoStop}%
\bibitem [{\citenamefont {Liu}\ \emph {et~al.}(2025)\citenamefont {Liu}, \citenamefont {Singh}, \citenamefont {Kennedy}, \citenamefont {Huang}, \citenamefont {Fluckey}, \citenamefont {Matthews},\ and\ \citenamefont {Uberuaga}}]{Liu2025}%
  \BibitemOpen
  \bibfield  {author} {\bibinfo {author} {\bibfnamefont {X.-Y.}\ \bibnamefont {Liu}}, \bibinfo {author} {\bibfnamefont {C.~N.}\ \bibnamefont {Singh}}, \bibinfo {author} {\bibfnamefont {E.~R.}\ \bibnamefont {Kennedy}}, \bibinfo {author} {\bibfnamefont {S.}~\bibnamefont {Huang}}, \bibinfo {author} {\bibfnamefont {S.~P.}\ \bibnamefont {Fluckey}}, \bibinfo {author} {\bibfnamefont {C.}~\bibnamefont {Matthews}},\ and\ \bibinfo {author} {\bibfnamefont {B.~P.}\ \bibnamefont {Uberuaga}},\ }\bibinfo {title} {Review of radiation-induced defects in GaAs},\ \href {https://doi.org/10.1063/5.0279267} {\bibfield  {journal} {\bibinfo  {journal} {Journal of Applied Physics}\ }\textbf {\bibinfo {volume} {138}},\ \bibinfo {pages} {070701} (\bibinfo {year} {2025})}\BibitemShut {NoStop}%
\bibitem [{\citenamefont {Sooby}\ \emph {et~al.}(2025)\citenamefont {Sooby}, \citenamefont {Warren}, \citenamefont {Tesmer}, \citenamefont {Houlton}, \citenamefont {Parker}, \citenamefont {Voges}, \citenamefont {McIntyre},\ and\ \citenamefont {Wang}}]{Sooby2025}%
  \BibitemOpen
  \bibfield  {author} {\bibinfo {author} {\bibfnamefont {E.~S.}\ \bibnamefont {Sooby}}, \bibinfo {author} {\bibfnamefont {P.~H.}\ \bibnamefont {Warren}}, \bibinfo {author} {\bibfnamefont {J.}~\bibnamefont {Tesmer}}, \bibinfo {author} {\bibfnamefont {R.}~\bibnamefont {Houlton}}, \bibinfo {author} {\bibfnamefont {S.~S.}\ \bibnamefont {Parker}}, \bibinfo {author} {\bibfnamefont {M.}~\bibnamefont {Voges}}, \bibinfo {author} {\bibfnamefont {P.~M.}\ \bibnamefont {McIntyre}},\ and\ \bibinfo {author} {\bibfnamefont {Y.}~\bibnamefont {Wang}},\ }\bibinfo {title} {Coupled Proton Beam Irradiation and Molten Chloride Salt Corrosion of Nickel},\ \bibfield  {journal} {\bibinfo  {journal} {JOM}\ }\href {https://doi.org/10.1007/s11837-025-07791-4} {10.1007/s11837-025-07791-4} (\bibinfo {year} {2025})\BibitemShut {NoStop}%
\bibitem [{\citenamefont {Zhao}\ \emph {et~al.}(2025)\citenamefont {Zhao}, \citenamefont {Zhao}, \citenamefont {Wang}, \citenamefont {Xue}, \citenamefont {Yu},\ and\ \citenamefont {Xing}}]{Zhao2025}%
  \BibitemOpen
  \bibfield  {author} {\bibinfo {author} {\bibfnamefont {Q.}~\bibnamefont {Zhao}}, \bibinfo {author} {\bibfnamefont {Y.}~\bibnamefont {Zhao}}, \bibinfo {author} {\bibfnamefont {M.}~\bibnamefont {Wang}}, \bibinfo {author} {\bibfnamefont {S.}~\bibnamefont {Xue}}, \bibinfo {author} {\bibfnamefont {R.}~\bibnamefont {Yu}},\ and\ \bibinfo {author} {\bibfnamefont {W.}~\bibnamefont {Xing}},\ }\bibinfo {title} {Atomic Manipulation of Metal Oxide Heterointerfaces by Electron Beam Illumination},\ \href {https://doi.org/10.1021/acs.jpclett.5c00018} {\bibfield  {journal} {\bibinfo  {journal} {The Journal of Physical Chemistry Letters}\ }\textbf {\bibinfo {volume} {16}},\ \bibinfo {pages} {1865} (\bibinfo {year} {2025})},\ \Eprint {https://arxiv.org/abs/https://doi.org/10.1021/acs.jpclett.5c00018} {https://doi.org/10.1021/acs.jpclett.5c00018} \BibitemShut {NoStop}%
\bibitem [{\citenamefont {Li}\ \emph {et~al.}(2020)\citenamefont {Li}, \citenamefont {Macdonald}, \citenamefont {Yang}, \citenamefont {Qiu},\ and\ \citenamefont {Wang}}]{Li2020}%
  \BibitemOpen
  \bibfield  {author} {\bibinfo {author} {\bibfnamefont {Y.}~\bibnamefont {Li}}, \bibinfo {author} {\bibfnamefont {D.~D.}\ \bibnamefont {Macdonald}}, \bibinfo {author} {\bibfnamefont {J.}~\bibnamefont {Yang}}, \bibinfo {author} {\bibfnamefont {J.}~\bibnamefont {Qiu}},\ and\ \bibinfo {author} {\bibfnamefont {S.}~\bibnamefont {Wang}},\ }\bibinfo {title} {Point defect model for the corrosion of steels in supercritical water: Part I, film growth kinetics},\ \href {https://doi.org/https://doi.org/10.1016/j.corsci.2019.108280} {\bibfield  {journal} {\bibinfo  {journal} {Corrosion Science}\ }\textbf {\bibinfo {volume} {163}},\ \bibinfo {pages} {108280} (\bibinfo {year} {2020})}\BibitemShut {NoStop}%
\bibitem [{\citenamefont {Banerjee}\ \emph {et~al.}(2020)\citenamefont {Banerjee}, \citenamefont {Kohnert}, \citenamefont {Holby},\ and\ \citenamefont {Uberuaga}}]{Banerjee2020}%
  \BibitemOpen
  \bibfield  {author} {\bibinfo {author} {\bibfnamefont {A.}~\bibnamefont {Banerjee}}, \bibinfo {author} {\bibfnamefont {A.~A.}\ \bibnamefont {Kohnert}}, \bibinfo {author} {\bibfnamefont {E.~F.}\ \bibnamefont {Holby}},\ and\ \bibinfo {author} {\bibfnamefont {B.~P.}\ \bibnamefont {Uberuaga}},\ }\bibinfo {title} {Critical Assessment of the Thermodynamics of Vacancy Formation in Fe$_{2}$O$_{3}$ Using Hybrid Density Functional Theory},\ \href {https://doi.org/10.1021/acs.jpcc.0c07522} {\bibfield  {journal} {\bibinfo  {journal} {The Journal of Physical Chemistry C}\ }\textbf {\bibinfo {volume} {124}},\ \bibinfo {pages} {23988} (\bibinfo {year} {2020})},\ \Eprint {https://arxiv.org/abs/https://doi.org/10.1021/acs.jpcc.0c07522} {https://doi.org/10.1021/acs.jpcc.0c07522} \BibitemShut {NoStop}%
\bibitem [{\citenamefont {Banerjee}\ \emph {et~al.}(2021)\citenamefont {Banerjee}, \citenamefont {Kohnert}, \citenamefont {Holby},\ and\ \citenamefont {Uberuaga}}]{Banerjee2021}%
  \BibitemOpen
  \bibfield  {author} {\bibinfo {author} {\bibfnamefont {A.}~\bibnamefont {Banerjee}}, \bibinfo {author} {\bibfnamefont {A.~A.}\ \bibnamefont {Kohnert}}, \bibinfo {author} {\bibfnamefont {E.~F.}\ \bibnamefont {Holby}},\ and\ \bibinfo {author} {\bibfnamefont {B.~P.}\ \bibnamefont {Uberuaga}},\ }\bibinfo {title} {Interplay between defect transport and cation spin frustration in corundum-structured oxides},\ \href {https://doi.org/10.1103/PhysRevMaterials.5.034410} {\bibfield  {journal} {\bibinfo  {journal} {Physical Review Materials}\ }\textbf {\bibinfo {volume} {5}},\ \bibinfo {pages} {034410} (\bibinfo {year} {2021})}\BibitemShut {NoStop}%
\bibitem [{\citenamefont {Banerjee}\ \emph {et~al.}(2023)\citenamefont {Banerjee}, \citenamefont {Holby}, \citenamefont {Kohnert}, \citenamefont {Srivastava}, \citenamefont {Asta},\ and\ \citenamefont {Uberuaga}}]{Banerjee2023}%
  \BibitemOpen
  \bibfield  {author} {\bibinfo {author} {\bibfnamefont {A.}~\bibnamefont {Banerjee}}, \bibinfo {author} {\bibfnamefont {E.~F.}\ \bibnamefont {Holby}}, \bibinfo {author} {\bibfnamefont {A.~A.}\ \bibnamefont {Kohnert}}, \bibinfo {author} {\bibfnamefont {S.}~\bibnamefont {Srivastava}}, \bibinfo {author} {\bibfnamefont {M.}~\bibnamefont {Asta}},\ and\ \bibinfo {author} {\bibfnamefont {B.~P.}\ \bibnamefont {Uberuaga}},\ }\bibinfo {title} {Thermokinetics of point defects in $\alpha$-Fe$_{2}$O$_{3}$},\ \href {https://doi.org/10.1088/2516-1075/acd158} {\bibfield  {journal} {\bibinfo  {journal} {Electronic Structure}\ }\textbf {\bibinfo {volume} {5}},\ \bibinfo {pages} {024007} (\bibinfo {year} {2023})}\BibitemShut {NoStop}%
\bibitem [{\citenamefont {Hatton}\ and\ \citenamefont {Uberuaga}(2023)}]{Hatton2023}%
  \BibitemOpen
  \bibfield  {author} {\bibinfo {author} {\bibfnamefont {P.}~\bibnamefont {Hatton}}\ and\ \bibinfo {author} {\bibfnamefont {B.~P.}\ \bibnamefont {Uberuaga}},\ }\bibinfo {title} {Short range order in disordered spinels and the impact on cation vacancy transport},\ \href {https://doi.org/10.1039/D2TA06102C} {\bibfield  {journal} {\bibinfo  {journal} {Journal of Materials Chemistry A}\ }\textbf {\bibinfo {volume} {11}},\ \bibinfo {pages} {3471} (\bibinfo {year} {2023})}\BibitemShut {NoStop}%
\bibitem [{\citenamefont {Dudarev}\ \emph {et~al.}(1998)\citenamefont {Dudarev}, \citenamefont {Botton}, \citenamefont {Savrasov}, \citenamefont {Humphreys},\ and\ \citenamefont {Sutton}}]{Dudarev1998}%
  \BibitemOpen
  \bibfield  {author} {\bibinfo {author} {\bibfnamefont {S.~L.}\ \bibnamefont {Dudarev}}, \bibinfo {author} {\bibfnamefont {G.~A.}\ \bibnamefont {Botton}}, \bibinfo {author} {\bibfnamefont {S.~Y.}\ \bibnamefont {Savrasov}}, \bibinfo {author} {\bibfnamefont {C.~J.}\ \bibnamefont {Humphreys}},\ and\ \bibinfo {author} {\bibfnamefont {A.~P.}\ \bibnamefont {Sutton}},\ }\bibinfo {title} {Electron-energy-loss spectra and the structural stability of nickel oxide: An LSDA+U study},\ \href {https://doi.org/10.1103/PhysRevB.57.1505} {\bibfield  {journal} {\bibinfo  {journal} {Physical Review B}\ }\textbf {\bibinfo {volume} {57}},\ \bibinfo {pages} {1505} (\bibinfo {year} {1998})}\BibitemShut {NoStop}%
\bibitem [{\citenamefont {Mishra}\ and\ \citenamefont {Chun}(2015)}]{Mishra2015}%
  \BibitemOpen
  \bibfield  {author} {\bibinfo {author} {\bibfnamefont {M.}~\bibnamefont {Mishra}}\ and\ \bibinfo {author} {\bibfnamefont {D.-M.}\ \bibnamefont {Chun}},\ }\bibinfo {title} {$\alpha$-Fe$_{2}$O$_{3}$ as a photocatalytic material: A review},\ \href {https://doi.org/https://doi.org/10.1016/j.apcata.2015.03.023} {\bibfield  {journal} {\bibinfo  {journal} {Applied Catalysis A: General}\ }\textbf {\bibinfo {volume} {498}},\ \bibinfo {pages} {126} (\bibinfo {year} {2015})}\BibitemShut {NoStop}%
\bibitem [{\citenamefont {Cao}\ \emph {et~al.}(2006)\citenamefont {Cao}, \citenamefont {Qiu}, \citenamefont {Liang}, \citenamefont {Zhao},\ and\ \citenamefont {Zhu}}]{Cao2006}%
  \BibitemOpen
  \bibfield  {author} {\bibinfo {author} {\bibfnamefont {H.}~\bibnamefont {Cao}}, \bibinfo {author} {\bibfnamefont {X.}~\bibnamefont {Qiu}}, \bibinfo {author} {\bibfnamefont {Y.}~\bibnamefont {Liang}}, \bibinfo {author} {\bibfnamefont {M.}~\bibnamefont {Zhao}},\ and\ \bibinfo {author} {\bibfnamefont {Q.}~\bibnamefont {Zhu}},\ }\bibinfo {title} {Sol-gel synthesis and photoluminescence of p-type semiconductor Cr$_{2}$O$_{3}$ nanowires},\ \href {https://doi.org/10.1063/1.2213204} {\bibfield  {journal} {\bibinfo  {journal} {Applied Physics Letters}\ }\textbf {\bibinfo {volume} {88}},\ \bibinfo {pages} {241112} (\bibinfo {year} {2006})}\BibitemShut {NoStop}%
\bibitem [{\citenamefont {Abdullah}\ \emph {et~al.}(2014)\citenamefont {Abdullah}, \citenamefont {Rajab},\ and\ \citenamefont {Al-Abbas}}]{Abdullah2014}%
  \BibitemOpen
  \bibfield  {author} {\bibinfo {author} {\bibfnamefont {M.~M.}\ \bibnamefont {Abdullah}}, \bibinfo {author} {\bibfnamefont {F.~M.}\ \bibnamefont {Rajab}},\ and\ \bibinfo {author} {\bibfnamefont {S.~M.}\ \bibnamefont {Al-Abbas}},\ }\bibinfo {title} {Structural and optical characterization of Cr$_{2}$O$_{3}$ nanostructures: Evaluation of its dielectric properties},\ \href {https://doi.org/10.1063/1.4867012} {\bibfield  {journal} {\bibinfo  {journal} {AIP Advances}\ }\textbf {\bibinfo {volume} {4}},\ \bibinfo {pages} {027121} (\bibinfo {year} {2014})}\BibitemShut {NoStop}%
\bibitem [{\citenamefont {Lide}(2007)}]{Lide2007}%
  \BibitemOpen
  \bibfield  {author} {\bibinfo {author} {\bibfnamefont {D.~R.}\ \bibnamefont {Lide}},\ }\href@noop {} {\emph {\bibinfo {title} {CRC Handbook of Chemistry and Physics}}},\ \bibinfo {edition} {88th}\ ed.,\ edited by\ \bibinfo {editor} {\bibfnamefont {D.~R.}\ \bibnamefont {Lide}}\ (\bibinfo  {publisher} {CRC Press: Boca Raton, FL},\ \bibinfo {year} {2007})\ \bibinfo {note} {the data for Fe$_{2}$O$_{3}$ originates from Samokhvalov, A. A., Sov. Phys.-Solid State 3 (1962) 2613.}\BibitemShut {Stop}%
\bibitem [{\citenamefont {Fang}\ and\ \citenamefont {Brower}(1963)}]{Fang1963}%
  \BibitemOpen
  \bibfield  {author} {\bibinfo {author} {\bibfnamefont {P.~H.}\ \bibnamefont {Fang}}\ and\ \bibinfo {author} {\bibfnamefont {W.~S.}\ \bibnamefont {Brower}},\ }\bibinfo {title} {Dielectric Constant of Cr$_{2}$O$_{3}$ Crystals},\ \href {https://doi.org/10.1103/PhysRev.129.1561} {\bibfield  {journal} {\bibinfo  {journal} {Phys. Rev.}\ }\textbf {\bibinfo {volume} {129}},\ \bibinfo {pages} {1561} (\bibinfo {year} {1963})}\BibitemShut {NoStop}%
\bibitem [{\citenamefont {Kresse}\ and\ \citenamefont {Joubert}(1999)}]{Kresse1999}%
  \BibitemOpen
  \bibfield  {author} {\bibinfo {author} {\bibfnamefont {G.}~\bibnamefont {Kresse}}\ and\ \bibinfo {author} {\bibfnamefont {D.}~\bibnamefont {Joubert}},\ }\bibinfo {title} {From ultrasoft pseudopotentials to the projector augmented-wave method},\ \href {https://doi.org/10.1103/PhysRevB.59.1758} {\bibfield  {journal} {\bibinfo  {journal} {Physical Review B}\ }\textbf {\bibinfo {volume} {59}},\ \bibinfo {pages} {1758} (\bibinfo {year} {1999})}\BibitemShut {NoStop}%
\bibitem [{\citenamefont {Kresse}\ and\ \citenamefont {Furthm\"{u}ller}(1996{\natexlab{a}})}]{Kresse1996a}%
  \BibitemOpen
  \bibfield  {author} {\bibinfo {author} {\bibfnamefont {G.}~\bibnamefont {Kresse}}\ and\ \bibinfo {author} {\bibfnamefont {J.}~\bibnamefont {Furthm\"{u}ller}},\ }\bibinfo {title} {Efficiency of ab-initio total energy calculations for metals and semiconductors using a plane-wave basis set},\ \href {https://doi.org/https://doi.org/10.1016/0927-0256(96)00008-0} {\bibfield  {journal} {\bibinfo  {journal} {Computational Materials Science}\ }\textbf {\bibinfo {volume} {6}},\ \bibinfo {pages} {15} (\bibinfo {year} {1996}{\natexlab{a}})}\BibitemShut {NoStop}%
\bibitem [{\citenamefont {Kresse}\ and\ \citenamefont {Furthm\"{u}ller}(1996{\natexlab{b}})}]{Kresse1996b}%
  \BibitemOpen
  \bibfield  {author} {\bibinfo {author} {\bibfnamefont {G.}~\bibnamefont {Kresse}}\ and\ \bibinfo {author} {\bibfnamefont {J.}~\bibnamefont {Furthm\"{u}ller}},\ }\bibinfo {title} {Efficient iterative schemes for ab initio total-energy calculations using a plane-wave basis set},\ \href {https://doi.org/10.1103/PhysRevB.54.11169} {\bibfield  {journal} {\bibinfo  {journal} {Physical Review B}\ }\textbf {\bibinfo {volume} {54}},\ \bibinfo {pages} {11169} (\bibinfo {year} {1996}{\natexlab{b}})}\BibitemShut {NoStop}%
\bibitem [{\citenamefont {Perdew}\ \emph {et~al.}(1996)\citenamefont {Perdew}, \citenamefont {Burke},\ and\ \citenamefont {Ernzerhof}}]{Perdew1996}%
  \BibitemOpen
  \bibfield  {author} {\bibinfo {author} {\bibfnamefont {J.~P.}\ \bibnamefont {Perdew}}, \bibinfo {author} {\bibfnamefont {K.}~\bibnamefont {Burke}},\ and\ \bibinfo {author} {\bibfnamefont {M.}~\bibnamefont {Ernzerhof}},\ }\bibinfo {title} {Generalized Gradient Approximation Made Simple},\ \href {https://doi.org/10.1103/PhysRevLett.77.3865} {\bibfield  {journal} {\bibinfo  {journal} {Physical Review Letters}\ }\textbf {\bibinfo {volume} {77}},\ \bibinfo {pages} {3865} (\bibinfo {year} {1996})}\BibitemShut {NoStop}%
\bibitem [{\citenamefont {G\"{o}khan~\"{U}nl\"{u}}\ \emph {et~al.}(2019)\citenamefont {G\"{o}khan~\"{U}nl\"{u}}, \citenamefont {Burak~Kaynar}, \citenamefont {\c{S}im\c{s}ek}, \citenamefont {Tekg\"{u}l}, \citenamefont {Kalkan},\ and\ \citenamefont {\"{O}zcan}}]{Gokhan2019}%
  \BibitemOpen
  \bibfield  {author} {\bibinfo {author} {\bibfnamefont {C.}~\bibnamefont {G\"{o}khan~\"{U}nl\"{u}}}, \bibinfo {author} {\bibfnamefont {M.}~\bibnamefont {Burak~Kaynar}}, \bibinfo {author} {\bibfnamefont {T.}~\bibnamefont {\c{S}im\c{s}ek}}, \bibinfo {author} {\bibfnamefont {A.}~\bibnamefont {Tekg\"{u}l}}, \bibinfo {author} {\bibfnamefont {B.}~\bibnamefont {Kalkan}},\ and\ \bibinfo {author} {\bibfnamefont {c.}~\bibnamefont {\"{O}zcan}},\ }\bibinfo {title} {Structure and magnetic properties of (La$_{1−x}$Fe$_{x}$)FeO$_{3}$ (x=0, 0.25, 0.50) perovskite},\ \href {https://doi.org/https://doi.org/10.1016/j.jallcom.2019.01.047} {\bibfield  {journal} {\bibinfo  {journal} {Journal of Alloys and Compounds}\ }\textbf {\bibinfo {volume} {784}},\ \bibinfo {pages} {1198} (\bibinfo {year} {2019})}\BibitemShut {NoStop}%
\bibitem [{\citenamefont {Fabrykiewicz}\ \emph {et~al.}(2018)\citenamefont {Fabrykiewicz}, \citenamefont {Przenios\l{}o}, \citenamefont {Sosnowska},\ and\ \citenamefont {Fauth}}]{Fabrykiewicz2018}%
  \BibitemOpen
  \bibfield  {author} {\bibinfo {author} {\bibfnamefont {P.}~\bibnamefont {Fabrykiewicz}}, \bibinfo {author} {\bibfnamefont {R.}~\bibnamefont {Przenios\l{}o}}, \bibinfo {author} {\bibfnamefont {I.}~\bibnamefont {Sosnowska}},\ and\ \bibinfo {author} {\bibfnamefont {F.}~\bibnamefont {Fauth}},\ }\bibinfo {title} {Positive and negative monoclinic deformation of corundum-type trigonal crystal structures of M$_{2}$O$_{3}$ metal oxides},\ \href {https://doi.org/10.1107/S2052520618014968} {\bibfield  {journal} {\bibinfo  {journal} {Acta Crystallographica Section B}\ }\textbf {\bibinfo {volume} {74}},\ \bibinfo {pages} {660} (\bibinfo {year} {2018})}\BibitemShut {NoStop}%
\bibitem [{\citenamefont {Kaspar}\ \emph {et~al.}(2023)\citenamefont {Kaspar}, \citenamefont {Spurgeon}, \citenamefont {Yano}, \citenamefont {Matthews}, \citenamefont {Bowden}, \citenamefont {Ophus}, \citenamefont {Kim}, \citenamefont {Wang},\ and\ \citenamefont {Schreiber}}]{Kaspar2023}%
  \BibitemOpen
  \bibfield  {author} {\bibinfo {author} {\bibfnamefont {T.~C.}\ \bibnamefont {Kaspar}}, \bibinfo {author} {\bibfnamefont {S.~R.}\ \bibnamefont {Spurgeon}}, \bibinfo {author} {\bibfnamefont {K.~H.}\ \bibnamefont {Yano}}, \bibinfo {author} {\bibfnamefont {B.~E.}\ \bibnamefont {Matthews}}, \bibinfo {author} {\bibfnamefont {M.~E.}\ \bibnamefont {Bowden}}, \bibinfo {author} {\bibfnamefont {C.}~\bibnamefont {Ophus}}, \bibinfo {author} {\bibfnamefont {H.}~\bibnamefont {Kim}}, \bibinfo {author} {\bibfnamefont {Y.~Q.}\ \bibnamefont {Wang}},\ and\ \bibinfo {author} {\bibfnamefont {D.~K.}\ \bibnamefont {Schreiber}},\ }\bibinfo {title} {Role of structural defects in mediating disordering processes at irradiated epitaxial Fe$_{3}$O$_{4}$/Cr$_{2}$O$_{3}$ interfaces},\ \href {https://doi.org/10.1103/PhysRevMaterials.7.093604} {\bibfield  {journal} {\bibinfo  {journal} {Physical Review Materials}\ }\textbf {\bibinfo {volume} {7}},\ \bibinfo {pages} {093604} (\bibinfo {year} {2023})}\BibitemShut {NoStop}%
\bibitem [{\citenamefont {Ziegler}\ \emph {et~al.}(2013)\citenamefont {Ziegler}, \citenamefont {Biersack},\ and\ \citenamefont {Ziegler}}]{Ziegler2013}%
  \BibitemOpen
  \bibfield  {author} {\bibinfo {author} {\bibfnamefont {J.~F.}\ \bibnamefont {Ziegler}}, \bibinfo {author} {\bibfnamefont {J.~P.}\ \bibnamefont {Biersack}},\ and\ \bibinfo {author} {\bibfnamefont {M.~D.}\ \bibnamefont {Ziegler}},\ }\bibinfo {title} {Stopping Range of Ions in Matter (SRIM)},\ \bibfield  {journal} {\bibinfo  {journal} {Lulu Press Co.:Morrisville, NC}\ }\href {https://doi.org/10.1016/j.nimb.2010.02.091} {10.1016/j.nimb.2010.02.091} (\bibinfo {year} {2013})\BibitemShut {NoStop}%
\bibitem [{\citenamefont {Yuan}\ \emph {et~al.}(2019)\citenamefont {Yuan}, \citenamefont {Zhang}, \citenamefont {Saeedi}, \citenamefont {Cheng}, \citenamefont {Guo}, \citenamefont {Liao}, \citenamefont {Zhang}, \citenamefont {Ying},\ and\ \citenamefont {Gehring}}]{Yuan2019}%
  \BibitemOpen
  \bibfield  {author} {\bibinfo {author} {\bibfnamefont {M.}~\bibnamefont {Yuan}}, \bibinfo {author} {\bibfnamefont {X.}~\bibnamefont {Zhang}}, \bibinfo {author} {\bibfnamefont {A.~M.}\ \bibnamefont {Saeedi}}, \bibinfo {author} {\bibfnamefont {W.}~\bibnamefont {Cheng}}, \bibinfo {author} {\bibfnamefont {C.}~\bibnamefont {Guo}}, \bibinfo {author} {\bibfnamefont {B.}~\bibnamefont {Liao}}, \bibinfo {author} {\bibfnamefont {X.}~\bibnamefont {Zhang}}, \bibinfo {author} {\bibfnamefont {I.}~\bibnamefont {Ying}},\ and\ \bibinfo {author} {\bibfnamefont {G.~A.}\ \bibnamefont {Gehring}},\ }\bibinfo {title} {Study of the radiation damage caused by ion implantation in ZnO and its relation to magnetism},\ \href {https://doi.org/https://doi.org/10.1016/j.nimb.2019.06.013} {\bibfield  {journal} {\bibinfo  {journal} {Nuclear Instruments and Methods in Physics Research Section B: Beam Interactions with Materials and Atoms}\ }\textbf {\bibinfo {volume} {455}},\ \bibinfo {pages} {7} (\bibinfo {year} {2019})}\BibitemShut
  {NoStop}%
\bibitem [{\citenamefont {Beyer}\ \emph {et~al.}(2021)\citenamefont {Beyer}, \citenamefont {Munde}, \citenamefont {Firoozabadi}, \citenamefont {Heimes}, \citenamefont {Grieb}, \citenamefont {Rosenauer}, \citenamefont {M{\"u}ller-Caspary},\ and\ \citenamefont {Volz}}]{Beyer2021}%
  \BibitemOpen
  \bibfield  {author} {\bibinfo {author} {\bibfnamefont {A.}~\bibnamefont {Beyer}}, \bibinfo {author} {\bibfnamefont {M.~S.}\ \bibnamefont {Munde}}, \bibinfo {author} {\bibfnamefont {S.}~\bibnamefont {Firoozabadi}}, \bibinfo {author} {\bibfnamefont {D.}~\bibnamefont {Heimes}}, \bibinfo {author} {\bibfnamefont {T.}~\bibnamefont {Grieb}}, \bibinfo {author} {\bibfnamefont {A.}~\bibnamefont {Rosenauer}}, \bibinfo {author} {\bibfnamefont {K.}~\bibnamefont {M{\"u}ller-Caspary}},\ and\ \bibinfo {author} {\bibfnamefont {K.}~\bibnamefont {Volz}},\ }\bibinfo {title} {Quantitative Characterization of Nanometer-Scale Electric Fields via Momentum-Resolved STEM},\ \href {https://doi.org/10.1021/acs.nanolett.0c04544} {\bibfield  {journal} {\bibinfo  {journal} {Nano Letters}\ }\textbf {\bibinfo {volume} {21}},\ \bibinfo {pages} {2018} (\bibinfo {year} {2021})},\ \bibinfo {note} {pMID: 33621104},\ \Eprint {https://arxiv.org/abs/https://doi.org/10.1021/acs.nanolett.0c04544} {https://doi.org/10.1021/acs.nanolett.0c04544}
  \BibitemShut {NoStop}%
\bibitem [{\citenamefont {Toyama}\ \emph {et~al.}(2022)\citenamefont {Toyama}, \citenamefont {Seki}, \citenamefont {Kanitani}, \citenamefont {Kudo}, \citenamefont {Tomiya}, \citenamefont {Ikuhara},\ and\ \citenamefont {Shibata}}]{Toyama2022}%
  \BibitemOpen
  \bibfield  {author} {\bibinfo {author} {\bibfnamefont {S.}~\bibnamefont {Toyama}}, \bibinfo {author} {\bibfnamefont {T.}~\bibnamefont {Seki}}, \bibinfo {author} {\bibfnamefont {Y.}~\bibnamefont {Kanitani}}, \bibinfo {author} {\bibfnamefont {Y.}~\bibnamefont {Kudo}}, \bibinfo {author} {\bibfnamefont {S.}~\bibnamefont {Tomiya}}, \bibinfo {author} {\bibfnamefont {Y.}~\bibnamefont {Ikuhara}},\ and\ \bibinfo {author} {\bibfnamefont {N.}~\bibnamefont {Shibata}},\ }\bibinfo {title} {Quantitative electric field mapping in semiconductor heterostructures via tilt-scan averaged DPC STEM},\ \href {https://doi.org/https://doi.org/10.1016/j.ultramic.2022.113538} {\bibfield  {journal} {\bibinfo  {journal} {Ultramicroscopy}\ }\textbf {\bibinfo {volume} {238}},\ \bibinfo {pages} {113538} (\bibinfo {year} {2022})}\BibitemShut {NoStop}%
\bibitem [{\citenamefont {Chejarla}\ \emph {et~al.}(2023)\citenamefont {Chejarla}, \citenamefont {Ahmed}, \citenamefont {Belz}, \citenamefont {Scheunert}, \citenamefont {Beyer},\ and\ \citenamefont {Volz}}]{Chejarla2023}%
  \BibitemOpen
  \bibfield  {author} {\bibinfo {author} {\bibfnamefont {V.~S.}\ \bibnamefont {Chejarla}}, \bibinfo {author} {\bibfnamefont {S.}~\bibnamefont {Ahmed}}, \bibinfo {author} {\bibfnamefont {J.}~\bibnamefont {Belz}}, \bibinfo {author} {\bibfnamefont {J.}~\bibnamefont {Scheunert}}, \bibinfo {author} {\bibfnamefont {A.}~\bibnamefont {Beyer}},\ and\ \bibinfo {author} {\bibfnamefont {K.}~\bibnamefont {Volz}},\ }\bibinfo {title} {Measuring Spatially-Resolved Potential Drops at Semiconductor Hetero-Interfaces Using 4D-STEM},\ \href {https://doi.org/https://doi.org/10.1002/smtd.202300453} {\bibfield  {journal} {\bibinfo  {journal} {Small Methods}\ }\textbf {\bibinfo {volume} {7}},\ \bibinfo {pages} {2300453} (\bibinfo {year} {2023})},\ \Eprint {https://arxiv.org/abs/https://onlinelibrary.wiley.com/doi/pdf/10.1002/smtd.202300453} {https://onlinelibrary.wiley.com/doi/pdf/10.1002/smtd.202300453} \BibitemShut {NoStop}%
\bibitem [{\citenamefont {Powell}\ and\ \citenamefont {Jablonski}(2002)}]{Powell2002}%
  \BibitemOpen
  \bibfield  {author} {\bibinfo {author} {\bibfnamefont {C.}~\bibnamefont {Powell}}\ and\ \bibinfo {author} {\bibfnamefont {A.}~\bibnamefont {Jablonski}},\ }\bibinfo {title} {The NIST Electron Effective-Attenuation-Length Database},\ \href {https://doi.org/10.1384/jsa.9.322} {\bibfield  {journal} {\bibinfo  {journal} {Journal of Surface Analysis}\ }\textbf {\bibinfo {volume} {9}},\ \bibinfo {pages} {322} (\bibinfo {year} {2002})}\BibitemShut {NoStop}%
\bibitem [{\citenamefont {Egerton}(2011)}]{Egerton2011}%
  \BibitemOpen
  \bibfield  {author} {\bibinfo {author} {\bibfnamefont {R.~F.}\ \bibnamefont {Egerton}},\ }\href {https://link.springer.com/book/10.1007/978-1-4419-9583-4} {\emph {\bibinfo {title} {Electron Energy-Loss Spectroscopy in the Electron Microscope}}}\ (\bibinfo  {publisher} {Springer Science \& Business Media},\ \bibinfo {year} {2011})\BibitemShut {NoStop}%
\end{thebibliography}

\providecommand{\noopsort}[1]{}\providecommand{\singleletter}[1]{#1}%

\newpage\newpage
\begin{Large}
\begin{center}
Supplementary Information: Irradiation-induced amplification of electric fields at oxide interfaces as revealed by correlative DPC-STEM and DFT
\end{center}
\end{Large}

\setcounter{figure}{0}

\renewcommand{\thefigure}{S\arabic{figure}}

\begin{figure}[h]
    \centering
    \includegraphics[width=0.8\linewidth]{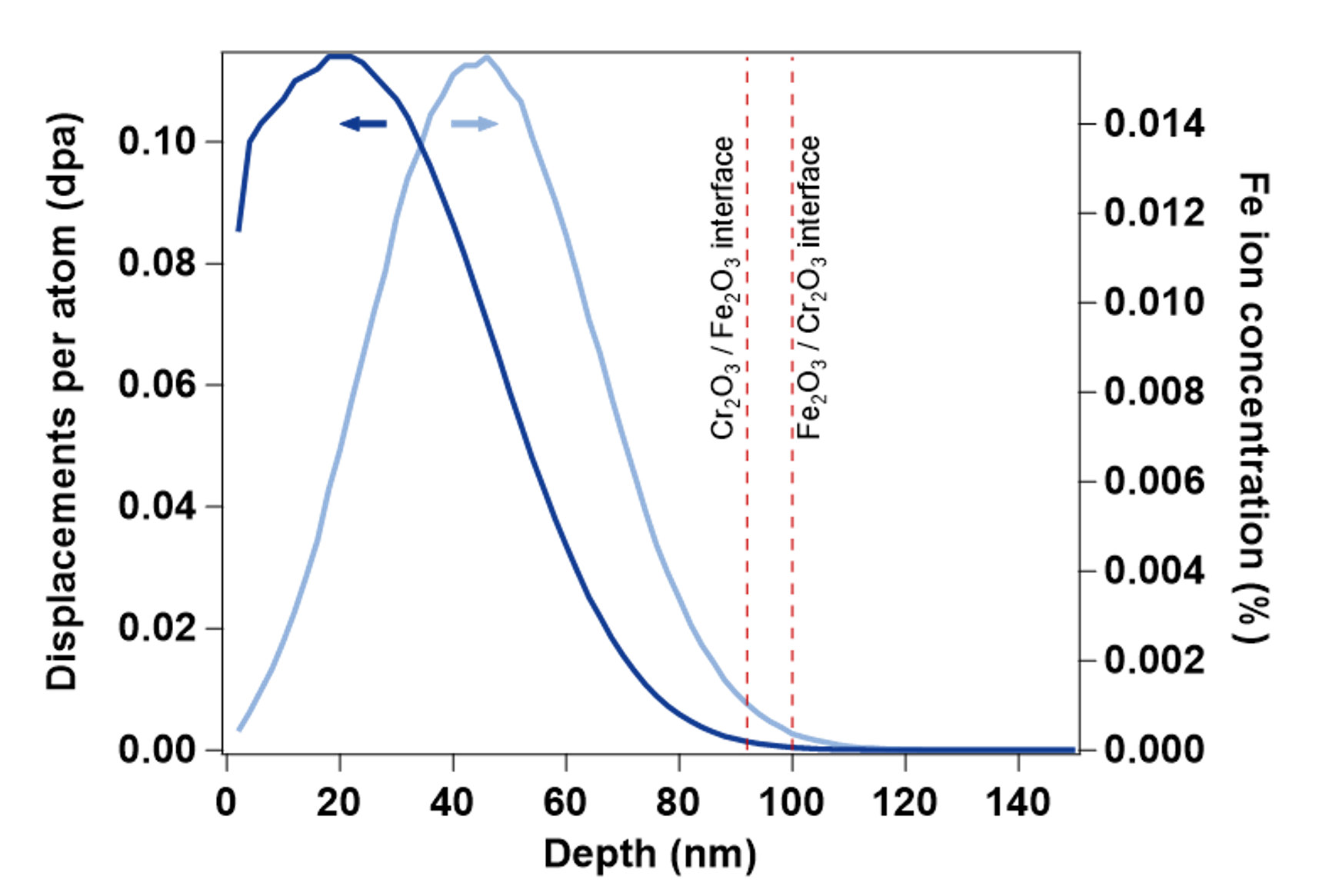}
    \caption{\textbf{The Stopping Range of Ions in Matter for the Fe+ Irradiation.} Stopping Range of Ions in Matter (SRIM) simulations in the Kinchin-Pease quick calculation mode of 100 keV Fe+ irradiation of 100 nm Fe$_{2}$O$_{3}$ / 100 nm Cr$_{2}$O$_{3}$ with the Fe$_{2}$O$_{3}$ film on top.  Displacement energies of Fe, Cr = 40 eV and O = 28 eV were used in the simulation. SRIM simulation results for 100 nm Cr$_{2}$O$_{3}$ / 100 nm Fe$_{2}$O$_{3}$ are very similar, therefore fluence values from Fe$_{2}$O$_{3}$ / Cr$_{2}$O$_{3}$ were used for the irradiation of both film stacks. Fluences were chosen to produce an average damage level of 0.1 dpa in the top 100 nm of the film stack (i.e., in the top film only). Dotted lines indicate the actual interface locations for both irradiated stacks, based on film thicknesses observed by cross-sectional STEM imaging.}
\end{figure}

\begin{figure}
    \centering
    \includegraphics[width=0.8\linewidth]{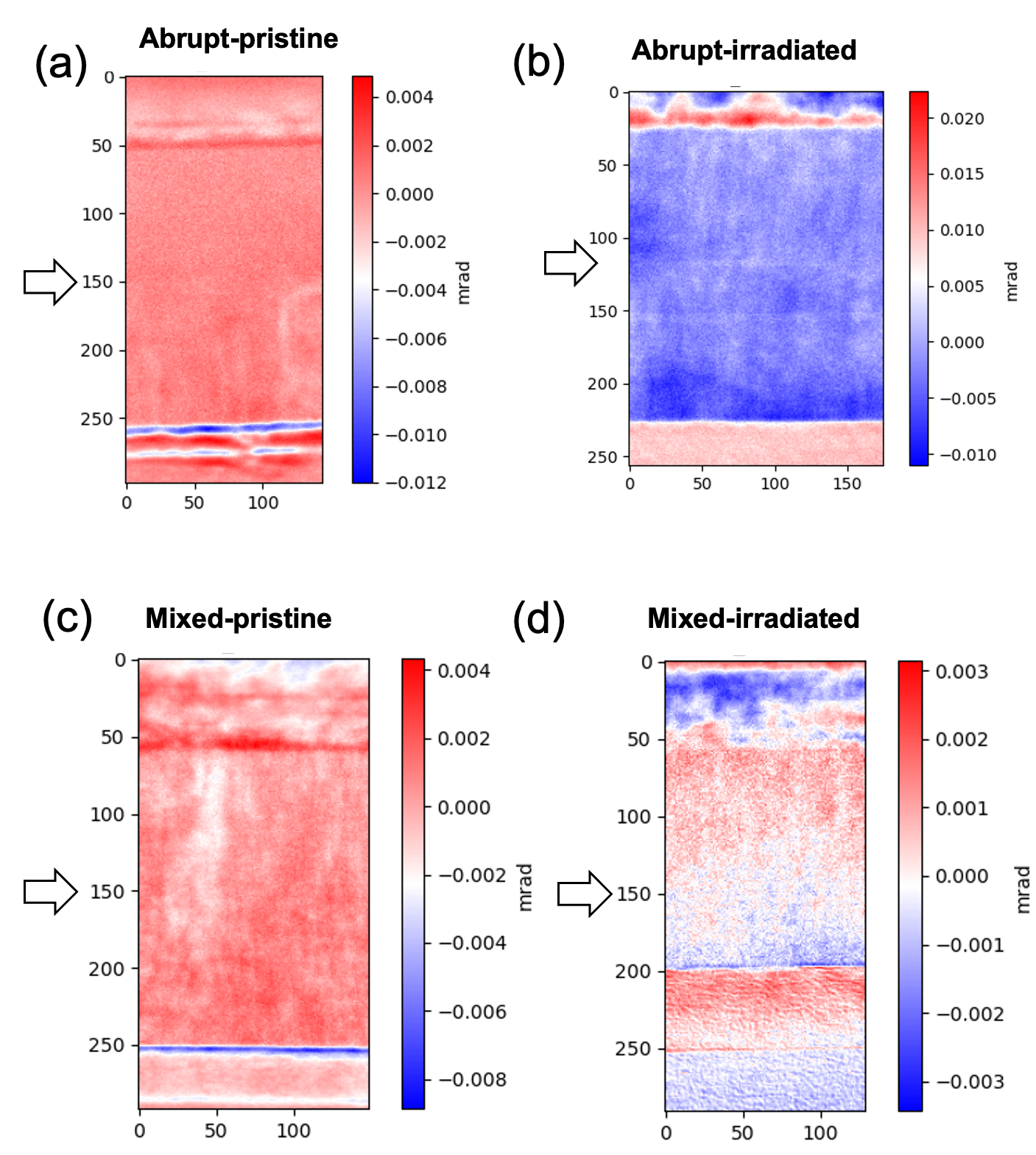}
    \caption{\textbf{Out-of-plane CoM shift (along the (0001)-axis) for all of the heterostructures.} Arrows indicate the location of the interface between Fe$_{2}$O$_{3}$ and Cr$_{2}$O$_{3}$ . The x and y axes are pixel indices. The size of each pixel is 1nm x 1 nm.}
\end{figure}

\begin{figure}
    \centering
    \includegraphics[width=0.8\linewidth]{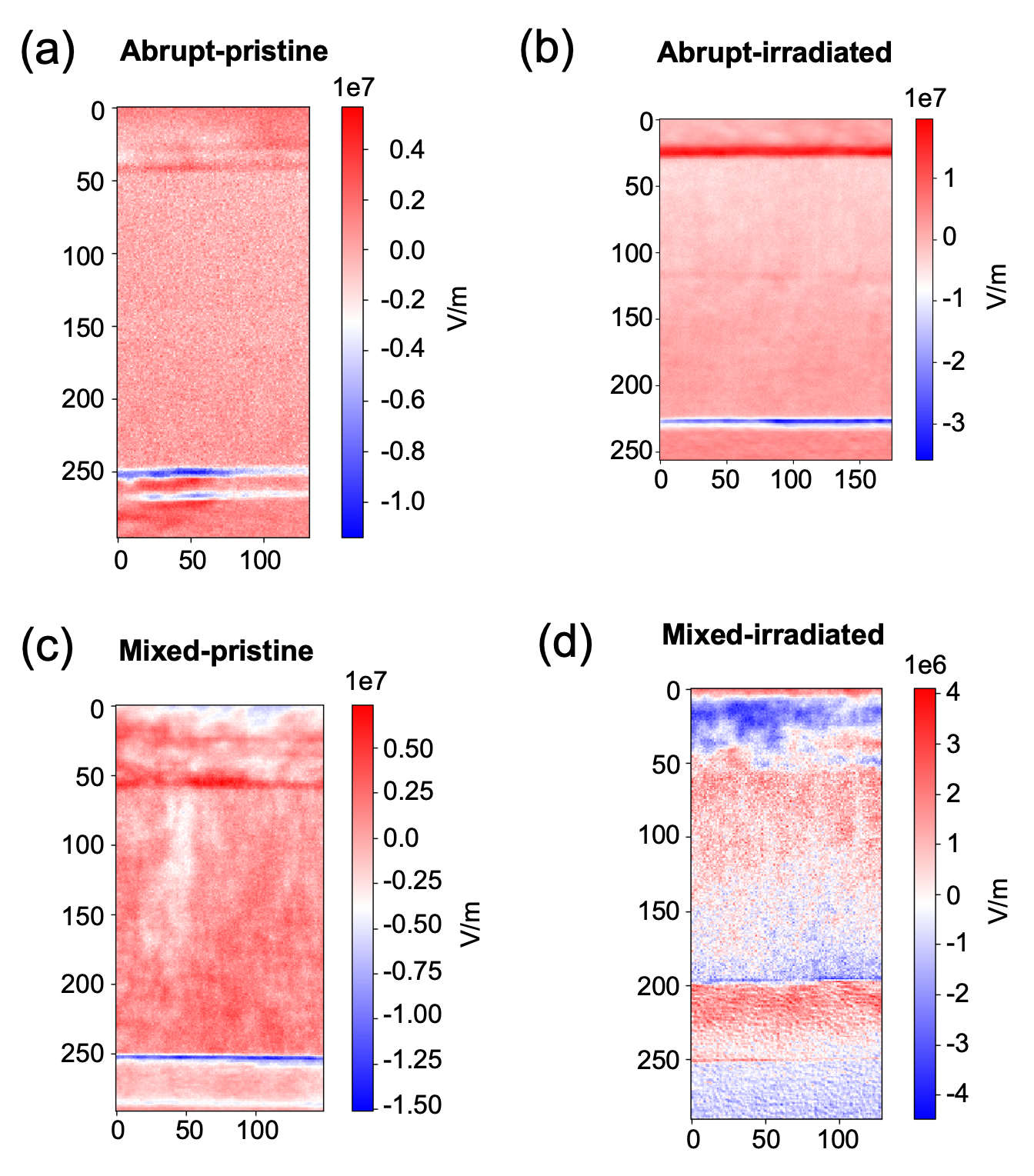}
    \caption{\textbf{Out-of-plane electric field (along the (0001)-axis) for all of the heterostructures.} The x and y axes are pixel indices. The size of each pixel is 1nm x 1 nm.}
\end{figure}

\begin{figure}
    \centering
    \includegraphics[width=0.8\linewidth]{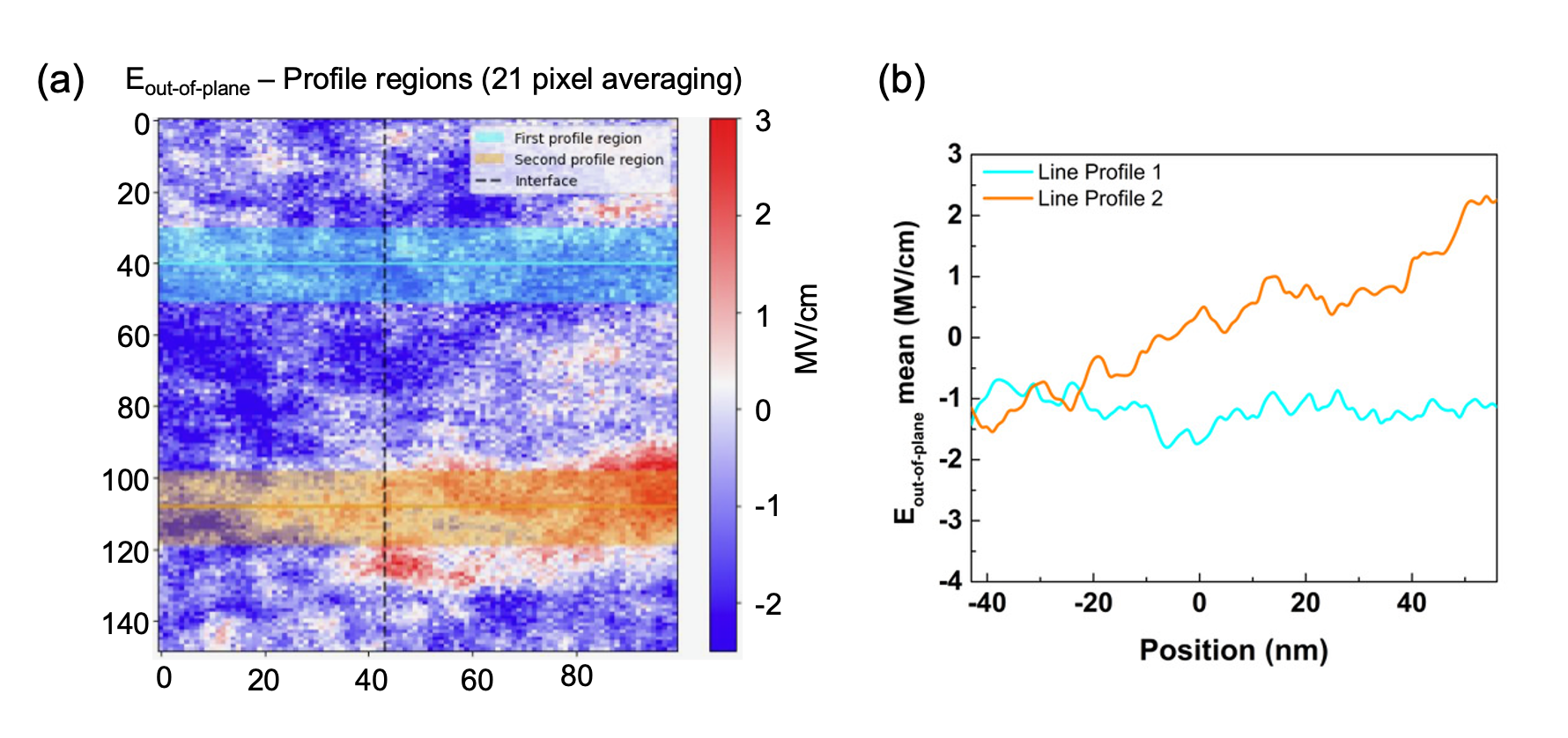}
    \caption{\textbf{Detailed out-of-plane electric field data for the mixed pristine interface.} (a) Out-of-plane electric field (along the (0001)-axis) for the mixed-pristine interface. The region highlighted in cyan is the region used in the main text to calculate the integrated potential and charge densities on either side of the interface. The region highlighted in orange is the region that experiences a highly localized change in out-of-plane electric field. The source of this localized variation from the rest of the sample is unknown. (b) The line profiles of the out-of-plane electric field for the cyan and orange regions.}
\end{figure}

\begin{figure}
    \centering
    \includegraphics[width=0.8\linewidth]{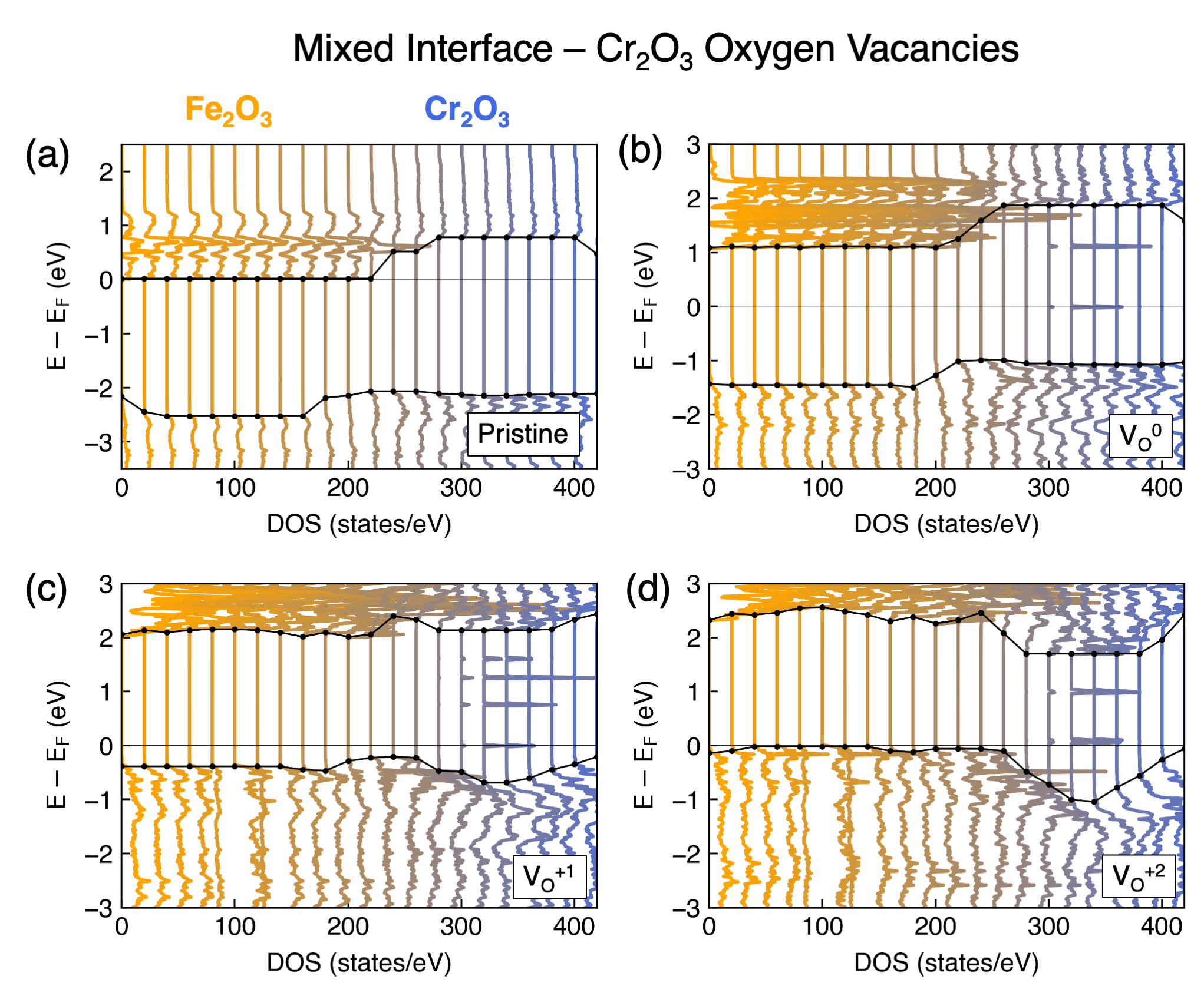}
    \caption{\textbf{The layer resolved density of states for the mixed interface for oxygen vacancies in Cr$_{2}$O$_{3}$.} Layer resolved density of states for the mixed interface in the (a) pristine, (b) neutral oxygen vacancy, (c) $+1$ charged oxygen vacancy, and (d) $+2$ charged oxygen vacancy states, for oxygen vacancies in the Cr$_{2}$O$_{3}$ layer. The oxygen vacancies introduce localized in-gap states. The bulk band edges are marked with black lines. The energies of the Cr$_{2}$O$_{3}$ band edges relative to the Fe$_{2}$O$_{3}$ band edges shift downward as the oxygen vacancy charge state becomes more positive. This should promote electron charge transfer from Fe$_{2}$O$_{3}$ to Cr$_{2}$O$_{3}$.}
\end{figure}

\begin{figure}
    \centering
    \includegraphics[width=0.8\linewidth]{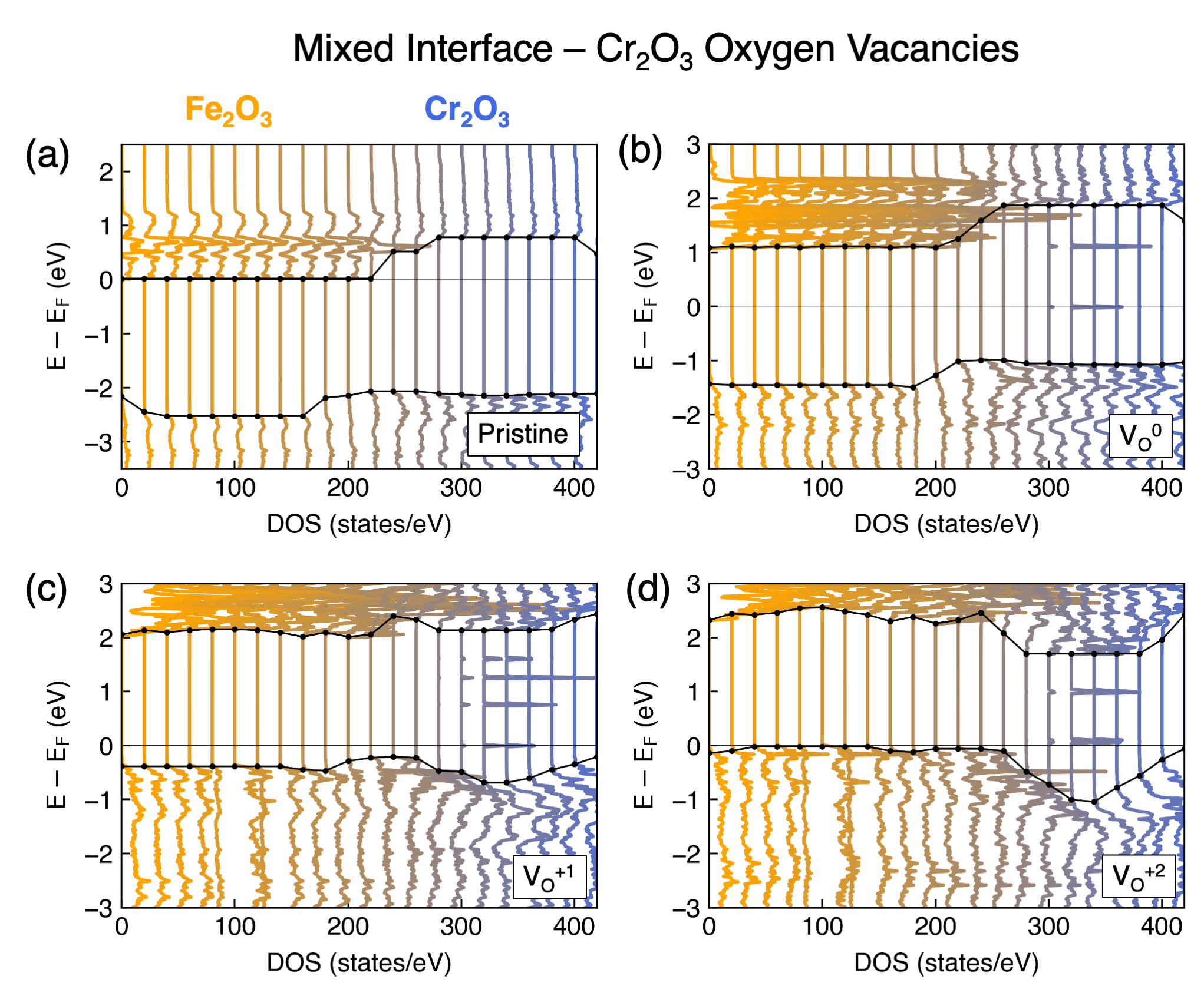}
    \caption{\textbf{The layer resolved density of states for the mixed interface for oxygen vacancies in Fe$_{2}$O$_{3}$.}  Layer resolved density of states for the abrupt interface in the (a) pristine, (b) neutral oxygen vacancy, (c) $+1$ charged oxygen vacancy, and (d) $+2$ charged oxygen vacancy states, for oxygen vacancies in the Fe$_{2}$O$_{3}$ layer. The oxygen vacancies introduce localized in-gap states. The bulk band edges are marked with black lines. The energies of the Cr$_{2}$O$_{3}$ band edges relative to the Fe$_{2}$O$_{3}$ band edges shift upward as the oxygen vacancy charge state becomes more positive. This should promote electron charge transfer from Cr$_{2}$O$_{3}$ to Fe$_{2}$O$_{3}$.}
\end{figure}

\begin{figure}
    \centering
    \includegraphics[width=0.8\linewidth]{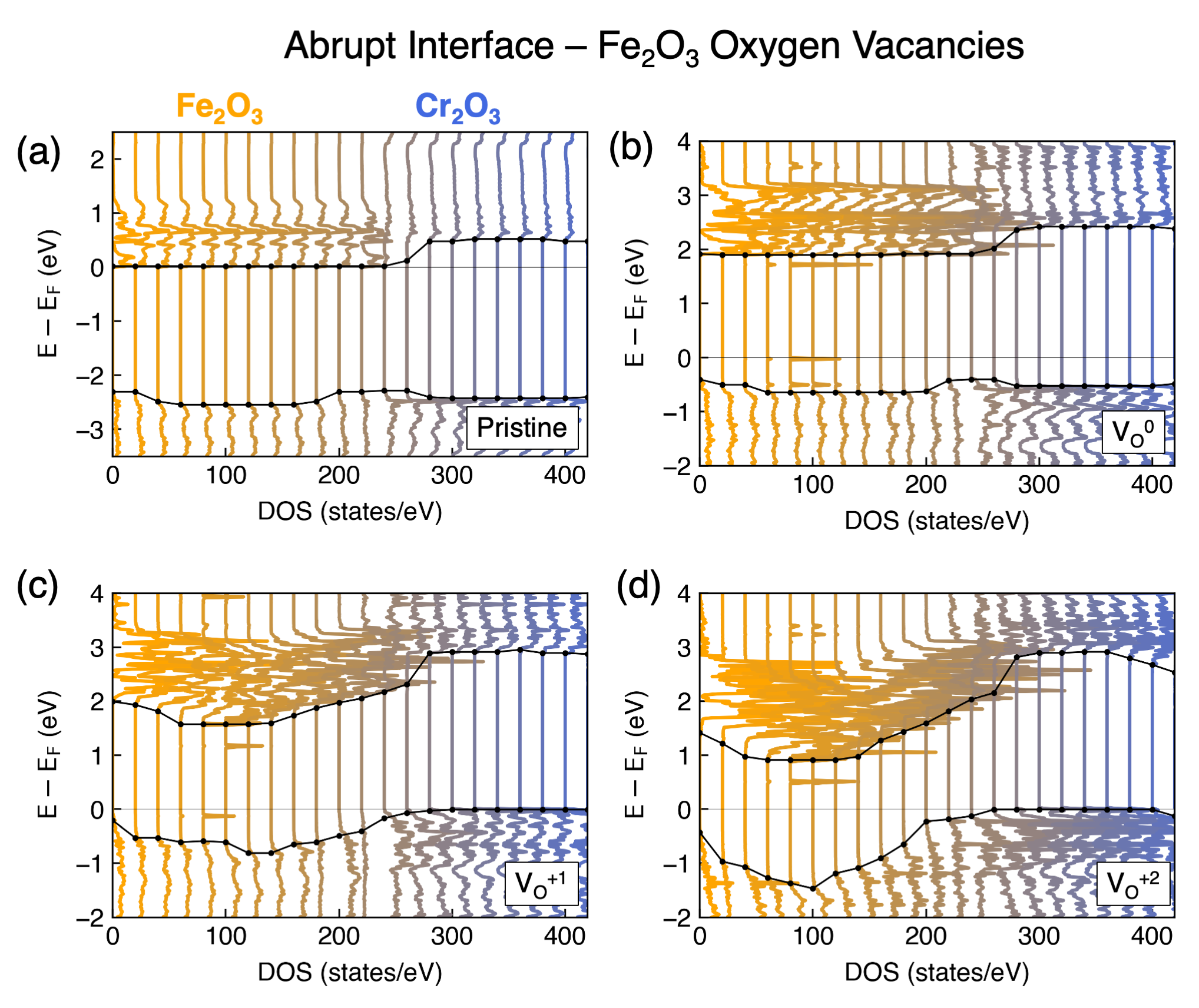}
    \caption{\textbf{The relative band offsets (neglecting in-gap defect states) of the abrupt and mixed interface heterostructures with oxygen vacancies immediately at the interface (within 0.14 nm).} For both the (a) abrupt interface with the Fe$_{2}$O$_{3}$ layer containing the oxygen vacancy and (b) the mixed interface with the Cr$_{2}$O$_{3}$ layer containing the oxygen vacancy, the band edges are largely unaffected by the presence of the oxygen vacancy, for all charge states.}
\end{figure}

\end{document}